\documentclass[prd,aps,a4paper,preprint,preprintnumbers,nofootinbib]{revtex4-1}



\usepackage[a4paper, hdivide={1.91cm,,1.165cm}, vdivide={1.83cm,,3.3cm}]{geometry}

\usepackage{amsmath,amssymb}
\usepackage{graphicx,multirow}
\usepackage{tabularx}
\usepackage[export]{adjustbox}
\usepackage[hyperfootnotes=false]{hyperref}
\usepackage[usenames, dvipsnames]{color}
\usepackage[normalem]{ulem}
\usepackage{float}

\newcommand{\dd}{\ensuremath{\text{d}}}

\newcommand{\beq}{\begin{eqnarray}}
\newcommand{\eeq}{\end{eqnarray}}

\begin{document}

\preprint{UCI-TR-2024-05}
\title{Mapping Excited Gauged Q-balls}

\author{Yahya Almumin\textsuperscript{1,2,}}
\email{yalmumin@uci.edu}

\affiliation{\textsuperscript{1}Department of Physics and Astronomy, University of California, Irvine, CA 92697-4575, USA}
\affiliation{\textsuperscript{2}Physics Department, Kuwait University, P.O. Box 5969 Safat, 13060, Kuwait}

\begin{abstract}
Properties such as the radius, charge and energy of ${U}(1)$ gauged Q-balls are analytically complicated to characterize. A mapping relation is known between the ground state of gauged Q-balls and global Q-balls that reduces the complexity of the analysis of the ground state of gauged Q-balls. Extending the map to excited states is a powerful tool to determine the properties of the rest of gauged Q-ball solution space. In this article we explore the mapping relation extension to characterize numerically and analytically excited gauged Q-balls solutions and their properties. A feature of excited gauged Q-balls is acquiring a maximum number excited states and having a maximal size, which we analytically predict via the mapping relation. We show that analytical approximations of the charge and energy, in the thin-wall limit, can be extended to excited gauged Q-balls and discuss the types of instabilities of these states.
\end{abstract}

\maketitle

\tableofcontents


\section{Introduction}
\label{I}

Global Q-balls are stable nontopological solitons composed of complex scalars that are charged under a global ${U}(1)$ and are held together by the scalar potential~\cite{Coleman:1985ki,Lee:1991ax}. The prominence of these configurations stems from their natural emergence in various particle physics models such as supersymmetry~\cite{Kusenko:1997zq,Enqvist:1997si}, extra dimensions~\cite{Demir:2000gj,Abel:2015tca}, and hidden sectors with QCD-like confinement~\cite{Bishara:2017otb,Bishara:2021fag}. The possible formation of Q-balls in early universe~\cite{Battye:2000qj,Multamaki:2002hv,Kusenko:2008zm,Tsumagari:2009na,Hiramatsu:2010dx} coupled with their stability~\cite{Coleman:1985ki} makes these configurations interesting dark matter candidates ~\cite{Kusenko:1997si,Kusenko:1997vp,Kusenko:2001vu,Ponton:2019hux,Bai:2019ogh,Bai:2021mzu}. Also, they have been discussed in the context of baryon asymmetry~\cite{Enqvist:1997si,Krylov:2013qe} and observable gravitational wave~\cite{Kusenko:2008zm,Croon:2019rqu}. To understand the full scope of global Q-balls, it is necessary to discuss excitations of these configurations. Excited states \cite{Volkov:2002aj,Mai:2012cx} will inevitably play a role in the formation and scattering of global Q-balls~\cite{Multamaki:2001az} despite their instability. Also, as pointed out in Ref.~\cite{Almumin_2022}, if these excited states are long-lived due to having small energy gap between these states and the ground, phenomenological implication could be relevant~\cite{Multamaki:2001az}. Even though we will concentrate in this article on radial excitations, charge-swapping Q-balls, which are  considered to be another type of excited Q-balls, have been studied extensively and shown to be qusai-stable~\cite{Copeland:2014qra,Xie:2021glp,Hou:2022jcd,Xie:2023psz}.

Solutions of global Q-balls are computed by solving a non-linear differential equation, which is solvable numerically for many cases~\cite{Masoumi:2016wot}. Analytically, some potentials are solvable exactly but they are usually unphysical~\cite{Rosen:1969ay,Theodorakis:2000bz,MacKenzie:2001av,Gulamov:2013ema}. For the sextic potential, where exact solutions are not attainable, excellent analytical approximation describing the ground state of global Q-balls have been produced in~\cite{Heeck:2020bau}. This paper was followed by Ref.~\cite{Almumin_2022} where the method have been extended to derive analytical approximations of properties of excited global Q-balls, in the thin-wall limit, for the same potential. Analytical approximations global Q-ball properties could be utilized in phenomenological studies of these configurations. 

 By promoting the ${U}(1)$ to a local symmetry, gauged Q-balls arise~\cite{Lee:1988ag} where we have a scalar and gauged fields coupled to one another. Gauged Q-balls have properties that are distinct from global Q-balls such as having a maximal possible size, and charge~\cite{Gulamov:2015fya,Brihaye:2015veu}. Studying these properties analytically by solving the two sets of coupled nonlinear equations in the gauged Q-ball set-up does not seem like a feasible strategy with the level of complexity each equation brings to the table. Luckily, a mapping relation between gauged and global Q-ball ground state was discovered in Ref.~\cite{Heeck_2021map}, which reduces the gauged Q-ball problem into essentially a global Q-ball one. Despite the differences between gauged and global Q-balls due to having the extra gauged field, the scalar field of the gauged Q-ball is still expected to admit radial excitations. Interestingly, radial excitations of gauged Q-balls in sextic potential have been studied in Ref.~\cite{Loginov_2020} where it was shown that excited states of gauged Q-balls also posses some unique properties compared to excited global Q-balls. For example, it was shown that the number of allowable excitations is finite for gauged Q-balls, whereas, it is infinite for global Q-balls.

In this article, we build on the success of the mapping relation in characterizing unexcited gauged Q-balls in terms of unexcited global Q-balls by extending the map to excited gauged Q-balls. By doing so, we find analytical expressions describing properties of these states such as the maximum possible number of excitation gauged Q-balls could acquire, and the maximum size of these Q-balls. We also demonstrate approximations of the charge and energy in the large  radius limit deduced from the mapping relation for the unexcited gauged Q-balls discovered in Ref.~\cite{Heeck_2021map} holds for excited gauged Q-ball and we comment on their instabilities. This implies that despite excited gauged Q-balls richer structure their properties are characterized by a single radius like length scale similar to the ground state. In our charge and energy discussion, we point out that in certain regions of the gauged Q-ball parameter space the energy gap between excited states and the ground state is  small, which suggest a longer lifetime for the excited states. Moreover, the numerical success of extending the mapping relation is also illustrated by using the finite element method to produce exact profiles of excited gauged Q-balls.

The article is organized as follows. We start the discussion in Sec.~\ref{II} by first reviewing unexcited and excited global Q-balls, unexcited gauged Q-balls, and how unexcited gauged and global Q-balls are related to one another via the mapping. In Sec.~\ref{III}, we show how extending the mapping relation provide an easy tool to obtain numerical profiles of excited gauged Q-balls. Analytical limits are derived for the maximum number of excitations a gauged Q-ball could have and the maximum possible radius for each excitation in Sec.~\ref{IV} via the mapping relation. Comparisons between the numerical and analytical prediction are shown in the same section. In Sec.~\ref{V}, we demonstrate the success of the charge and energy approximations in estimating the exact values of these quantities in the excited gauged Q-ball case in the large radius limit. We also point out that for certain regions of the parameter space excited gauged Q-balls could have longer lifetime due to the small energy gap with the ground stand and discuss different instability regions. Our summary and conclusion are presented in Sec.~\ref{sum}.

\section{Review of Q-balls}
\label{II}
Preliminary concepts about global and gauged Q-balls will aid the discussion about excited gauged Q-balls in light of excited global Q-balls properties. Therefore, a recap of global Q-balls and gauged Q-balls is presented in this section. 

\subsection{Global Q-balls}
Let us look at a general Lagrange density that produces global Q-ball solutions
\begin{equation}
\mathcal{L}=| \partial_\mu \phi| -U(| \phi |)\,,\label{1}
\end{equation}
where we have a complex scalar field $\phi$ and a potential $U(|\phi|)$. The Lagrangian is symmetric under an unbroken global ${U}(1)$, which implies that in the vacuum $\langle\phi\rangle=0$. We normalize our potential to be zero at the vacuum $U(0)=0$ and impose that the vacuum is a stable minimum 
\begin{equation}
  \frac{dU}{d|\phi|}=0\,, \ \ \ \frac{d^2U}{d\phi d\phi^*}|_{\phi=0}= m^2_\phi>0\,,\label{2}
\end{equation}
where $m_{\phi}$ is the mass of the complex scalar field. Coleman showed \cite{Coleman:1985ki} that for a potential to support the existence of a non-topological Q-ball soliton, the function $U(|\phi|)/|\phi|^2$ must have a minimum at $|\phi|=\phi_0/\sqrt{2}>0$ such that 
\begin{equation}
    0\leq \frac{\sqrt{2U(\phi_0/\sqrt{2})}}{\phi_0} \equiv \omega_0\leq m_0 \,.\label{3}
\end{equation}
These conditions will give rise to the simplest global Q-ball solution, which can be expressed in terms of a spherically symmetric solution.
\begin{equation}
    \phi(t,\vec{x})=\frac{\phi_0}{\sqrt{2}} f(r)e^{i\omega_G t}\,,
    \label{4}
\end{equation}
where $f(r)$ is a dimensionless time-independent function of radius $r$, and $\omega_G$ is the frequency bounded by $\omega_0<\omega_G<m_\phi$. Re-writing our parameters in terms of more traceable quantities that are dimensionless
\begin{equation}
    \rho\equiv r\sqrt{m_\phi ^2-\omega_0^2}\,,\ \ \ \Omega_G\equiv \frac{\omega_G}{\sqrt{m_\phi ^2-\omega_0^2}}\,, \ \ \ \Omega_0\equiv\frac{\omega_0}{\sqrt{m_\phi ^2-\omega_0^2}}\,, \ \ \ \Phi\equiv\frac{\phi_0}{\sqrt{m_\phi ^2-\omega_0^2}}\,.\label{5}
\end{equation}
The Lagrange density in terms of the $f(\rho)$ and dimensionless radius $\rho$ would be 
\begin{equation}
    \mathcal{L}=4\pi \Phi_0^2 \sqrt{m_\phi ^2-\omega_0^2}\int  d\rho \rho^2 \left[-\frac{1}{2}f'^2+\frac{1}{2}f^2\Omega_G^2-\frac{U(f)}{\Phi_0^2(m_\phi ^2-\omega_0^2)^2}\right]\,,\label{6}
\end{equation}
where the derivatives are in terms of $\rho$. The equation of motion defining the global Q-ball would be
\begin{equation}
     f''+\frac{2}{\rho}f'=\frac{1}{\Phi_0^2(m_\phi-\omega_0)^2}\frac{dU}{df}-\Omega_G^2f \,,\label{7}
\end{equation}
with the following boundary condition
\begin{equation}
   \lim_{\rho\to0}f'= \lim_{\rho\to\infty}f=0\,.
   \label{8}
\end{equation}
Throughout the discussion we will concentrate on the sextic potential that respects ${U}(1)$ symmetry 
\begin{equation}
    U(f)=\phi_0^2\left(\frac{m_\phi^2-\omega_0^2}{2}f^2\left(1-f^2\right)^2+\frac{\omega_0^2}{2}f^2\right).
\label{9}
\end{equation}
To understand the global Q-ball solution, we re-write the equation in terms of the effective
\begin{equation}
    f''+\frac{2}{\rho}f'+\frac{dV(f)}{df}=0\,,
\label{10}
\end{equation}
where the effective potential is defined as
\begin{equation}
    V(f)=\frac{1}{2}f^2\Omega^2-\frac{U(f)}{\phi_0^2}=\frac{1}{2}f^2\left[\kappa_G^2-\left(1-f^2\right)^2\right],
\label{11}
\end{equation}
where $\kappa_G^2\equiv\Omega_G^2-\Omega_0^2$ and bounded by $0<\kappa_G<1$. 

\begin{figure}[t]
\includegraphics[width=0.49\textwidth]{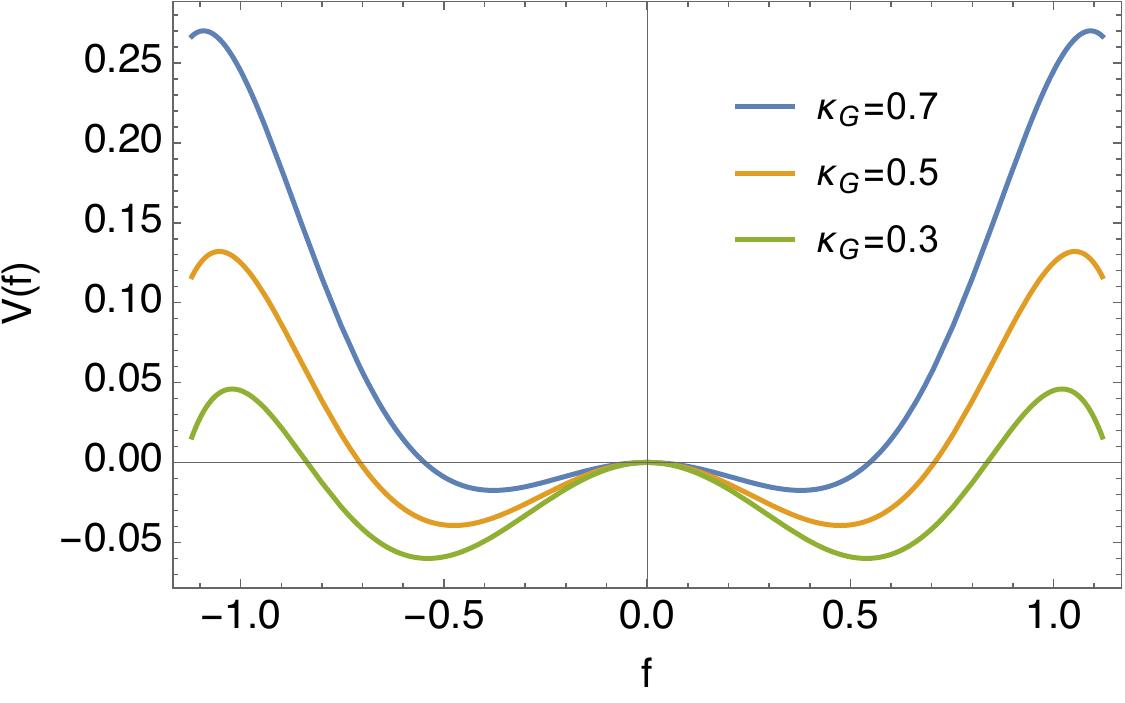}
\includegraphics[width=0.47\textwidth]{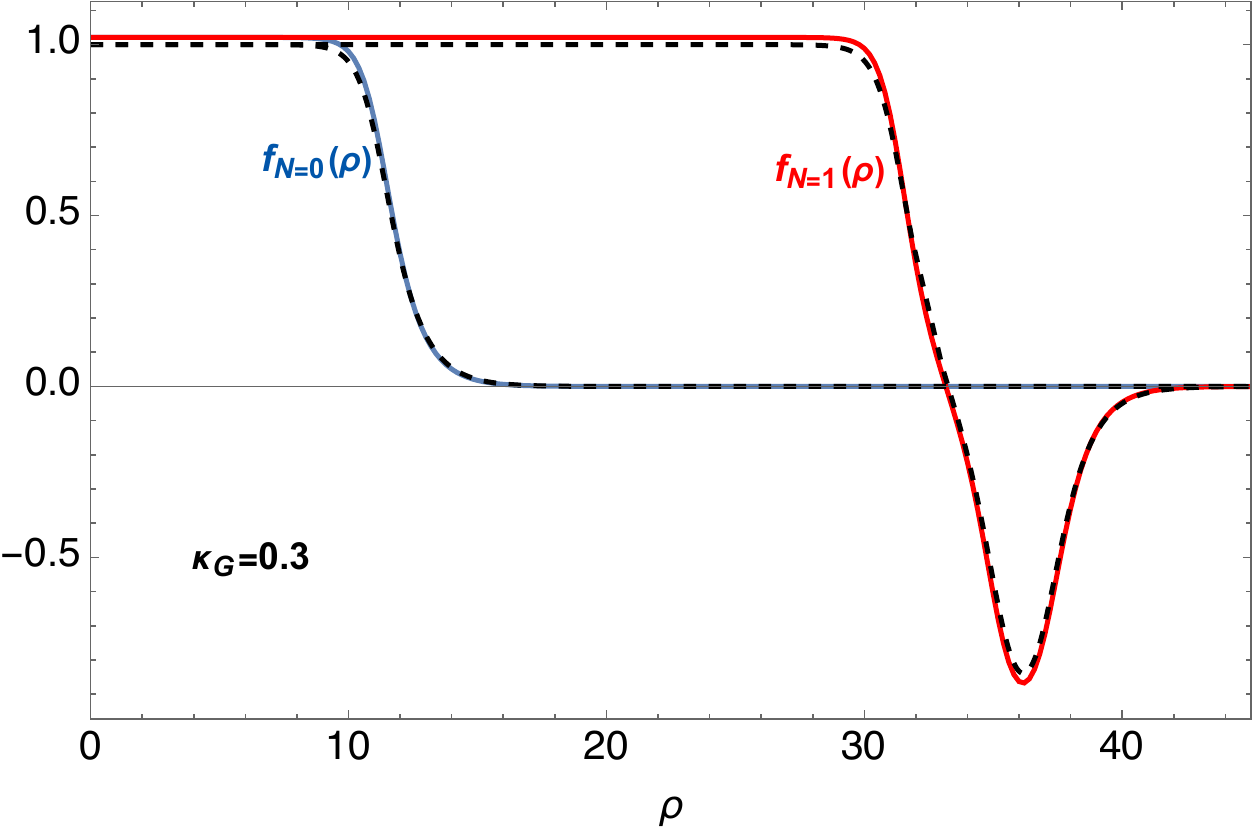}
\caption{Effective potentials of global Q-balls for different $\kappa_G$ vs $f$ (left). Exact profiles of the ground and first excited state of global Q-ball with $\kappa_G=0.3$ denoted by the solid lines and the global Q-ball that maps to the specified gauged Q-ball denoted by the dashed line. (right)
 }
\label{fig:1}
\end{figure}

The equation of motion (Eq.(\ref{10})) is analogous to a particle rolling in a potential with time-dependent friction~\cite{Coleman:1985ki} if we interpret the profile $f(\rho)$ as the position of the particle and dimensionless radius $\rho$ as the time the particle takes to move in potential. The particle starts at rest somewhere close to the top of the hill due to large friction and eventually transitions to $f=0$. The shape of the effective, which depends on $\kappa_G$, as shown in Fig.~\ref{fig:1}, determines the amount of time the particle rests on top before transitioning to zero i.e. determining the global Q-ball profile. The equation of motion can be satisfied in a number of ways since the particle can roll $N$ times past the $f=0$ until the particle comes to halt at $f=0$. The particle transitioning to zero directly represents the ground state of the global Q-ball solution, while the $Nth$ excited state ($N$ is an integer) are represented by the transition of the particle $N$ number of times past the $f=0$~\cite{Volkov:2002aj,Mai:2012cx}. 

In the thin wall limit ($\kappa_G$ is small), the ground state of global Q-ball radius that is defined as  $f''(R^*)=0$, is approximated by~\cite{Heeck:2020bau} 
\begin{equation}
    R^*_0(\kappa)=\frac{1}{\kappa_G^2}\,.
    \label{12}
\end{equation}
For the $Nth$ excited state of a global Q-ball we expect $\left(2N+1\right)$ radii $R^*_{N,n}$ satisfying $f''(R^*_{N,n})=0$ where $n$ is an integer that goes from $1$ to $N$. However, the length scale that characterize excited Q-balls properties is approximated by
\begin{equation}
    R^*_N(\kappa)=\frac{\left(2N+1\right)}{\kappa_G^2}\,,
    \label{60}
\end{equation}
where the $(2N+1)$ radii are expected to cluster~\cite{Almumin_2022}. The position of this quantity is not well defined in terms of the profile but the Q-ball properties and the conventional radii $R^*_{N,n}$ are derived from this length scale as shown in Ref.~\cite{Almumin_2022}. Throughout the discussion of the article when talking about the excited Q-ball radius we will refer to this characterizing length scale even though it does not satisfy the radius definition of the ground state.

A useful expression to  approximate the ground state profile of Q-balls would be the transition function 
\begin{equation}
    f_T=\frac{1}{\sqrt{1+2e^{2\left(\rho-R_0^*\right)}}}\,,
    \label{13}
\end{equation}
derived in Ref.~\cite{Heeck:2020bau}. In the thin-wall limit, the profile of the $Nth$ excited state of a global Q-ball, as shown in Fig.~\ref{fig:1}, can be approximated by the product of the transition functions~\cite{Almumin_2022}
\begin{align}
    f_N= \left[f_T(\rho,R^\ast_{N,1})-f_T(-\rho,-R^\ast_{N,2}) \right]\cdots\left[f_T(\rho,R^\ast_{N,2N-1})-f_T(-\rho,-R^\ast_{N,2N})\right]f_T(\rho,R^\ast_{N,2N+1}).
    \label{14}
\end{align}

Observable such as the charge $Q_G$ and energy $E_G$ of global Q-balls are expressed in terms of the following integrals
\begin{equation}
    Q_G=\frac{4\pi  \phi_0^2  \omega_G}{(m_\phi^2-\omega_0^2)^{3/2}} \int \dd\rho\,\rho^2f^2\,,
    \label{15}
\end{equation}
\begin{equation}
    E_G=\omega Q_G+\frac{4\pi\phi_0^2}{3\sqrt{m_\phi^2-\omega_0^2}}\int \dd\rho\,\rho^2f^{\prime 2}\,,
    \label{16}
\end{equation}
satisfying the the Q-ball condition $dE_G/d\omega=\omega dQ_G/d\omega$~\cite{Friedberg:1976me,Lee:1991ax,Heeck:2020bau}. The integrals are approximated to be~\cite{Almumin_2022}
\begin{equation}
   \int \dd\rho\,\rho^2f^2 \simeq \frac{(2N+1)^3}{3\kappa_G^6}\,,
   \label{17}
\end{equation}
\begin{equation}
 \int \dd\rho\,\rho^2f^{\prime 2}\simeq \frac{(2N+1)^3}{4\kappa_G^4}\,,
 \label{18}
 \end{equation}
allowing us to estimate the observables in terms of $\kappa_G$ for all excited states of a given Q-ball in the thin wall (large radius) limit.

\subsection{Gauged Q-balls}
Promoting the ${U}(1)$ to a local symmetry gives rise to gauged Q-balls. The Lagrange density for gauged Q-balls is 
\begin{equation}
    \mathcal{L}=|D_\mu\phi|^2-U(|\phi|)-\frac{1}{4} F_{\mu\nu}F^{\mu\nu},
    \label{19}
\end{equation}
where $D_\mu=\partial_\mu-ieA_\mu$ is the covariant derivative, $F_{\mu\nu}=\partial_\mu A_\nu -\partial_\nu A_\mu$ is the field-strength tensor, and $e$ is the normalized gauge coupling to insure $\phi$ has charge one.  The scalar and gauged field can be written following the static charge ansatz from Ref.~\cite{Lee:1988ag} as
\begin{equation}
    \phi(t,\vec{x})=\frac{\phi_0}{\sqrt{2}} f(r)e^{i\omega t},\ \ \ A_0(t,\vec{x})\equiv A_0(r), \ \ \ A_i(t,\vec{x})=0 \,.
    \label{20}
\end{equation}
The scalar frequency of gauged Q-balls is bounded by $\omega_0<\omega \leq m_\phi$, which differs slightly from the global case where $\omega \neq m_\phi$~\cite{Gulamov:2015fya}. Re-defining our parameters in terms of dimensionless quantities
\begin{equation}
    A(\rho)\equiv \frac{A_0(\rho)}{\phi_0}\,,\ \ \ \Omega\equiv \frac{\omega}{\sqrt{m_\phi ^2-\omega_0^2}}\,, \ \ \ \alpha\equiv e\Phi_0\,, \ \ \ \kappa \equiv\Omega^2-\Omega_0^2\,,
    \label{21}
\end{equation}
giving us the following Lagrangian 
\begin{equation}
    \mathcal{L}=4\pi \Phi_0^2 \sqrt{m_\phi ^2-\omega_0^2}\int  d\rho \rho^2 \left[-\frac{1}{2}f'^2+\frac{1}{2}A'^2+\frac{1}{2}f^2(\Omega^2-\alpha A)^2-\frac{U(f)}{\Phi_0^2(m_\phi ^2-\omega_0^2)^2}\right],
    \label{22}
\end{equation}
with the effective potential being
\beq
 V(f,A)=\frac{1}{2}f^2\left(\Omega-\alpha A \right)^2-\frac{U(f)}{\Phi_0^2(m_\phi^2-\omega_0^2)^2} =\frac{1}{2}f^2\left[\kappa^2+\alpha A(\alpha A-2\Omega)-\left(1-f^2\right)^2 \right],
 \label{23}
 \eeq
where we have expressed $V(f,A)$ in terms of the sextic potential. The effective potential in the the gauged Q-ball case depends on two fields $f$ and $A$ producing two equations of motions 
\begin{align}
f'' + \frac{2}{\rho} f'&= -\frac{\partial V}{\partial f} = \frac{1}{\Phi_0^2(m_\phi^2-\omega_0^2)^2}\frac{\dd U}
{\dd f}-\left(\Omega-\alpha A\right)^2f\,,\label{e.feq}\\
A'' + \frac{2}{\rho} A' &= +\frac{\partial V}{\partial A} = \alpha f^2(A\alpha-\Omega)\,,\label{24}
\end{align}
with the boundary condition being
\begin{equation}
\lim_{\rho\to0}f'= \lim_{\rho\to\infty}f =  \lim_{\rho\to0}A'= \lim_{\rho\to\infty}A =0 \,.
\label{26}
\end{equation}

Notice that in that the kinetic term of $A$ carries the opposite sign. This introduces a subtle distinction between the scalar field $f$ and the gauge field $A$ analysis in terms of the effective potential. Contrary to the scalar field, the gauge field $A$ is expected to start at somewhere downhill on the effective potential and get pushed upward toward the $A(\rho\rightarrow\infty) =0$. The effective potential has a minimum in terms of the gauge field when $f$ is constant at
\begin{equation}
    A_{\text{max}}=\frac{\Omega}{\alpha}\,,
    \label{27}
\end{equation}
and depending on the initial value of $A$, the gauged field would either go to zero or infinity for $\rho \rightarrow\infty$. Therefore, we expect that the initial $\Omega-\alpha A>0$ to satisfy the boundary condition for gauged Q-balls to exist, otherwise the gauge field will go to infinity~\cite{Heeck_2021map}. 

On the other hand the scalar field $f$ equation of motion in the gauged set-up, similar to the global case, is analogous to a particle rolling in the potential with time-dependent friction with the caveat that the $V(f,A)$ is dynamical in $\rho$ since $A$ changes with the radius (Fig.~\ref{2}). The particle still can roll $N$ time past $f=0$ before ending up there satisfying the boundary condition, which gives rise to excited gauged Q-ball solutions. The effective potential has three extrema when $A$ is constant one is at $f=0$ and the other two are at
\beq
f^2_\pm=\frac13\left(2\pm\sqrt{1+3\kappa^2-3\alpha A(2\Omega-\alpha A)} \right) ,\label{28}
\eeq
where $f_+$ is the maximum and $f_-$ is the minimum. The extrema introduces a boundary condition on $\alpha A$ as shown in~\cite{Heeck_2021map} 
\begin{equation}
    \alpha A \leq \Omega-\sqrt{\Omega_0^2-1/3}\,,
    \label{29}
\end{equation}
when $\Omega_0\leq1/\sqrt{3}$. The other boundary $\alpha A \geq \Omega-\sqrt{\Omega_0^2-1/3}$ is excluded since it violates the gauged Q-ball condition $\Omega-\alpha A>0$ as illustrated earlier. Observables of gauged Q-balls such as charge and energy~\cite{Lee:1988ag,Heeck_2021map} are expressed as 
\begin{equation}
    Q=\frac{4\pi\phi_0^2}{(m_\phi^2-\omega_0^2)} \int \dd\rho\,\rho^2f^2\left(\Omega-\alpha A \right),
    \label{30}
\end{equation}
\begin{equation}
    E=\omega Q+\frac{4\pi\phi_0^2}{3\sqrt{m_\phi^2-\omega_0^2}}\int \dd\rho\,\rho^2\left(f^{\prime 2}-A'^2\right),
    \label{31}
\end{equation}
satisfying $dE/d\omega=\omega dQ/d\omega$~\cite{Gulamov:2013cra}  .

\begin{figure}[t]
\includegraphics[width=0.485\textwidth]{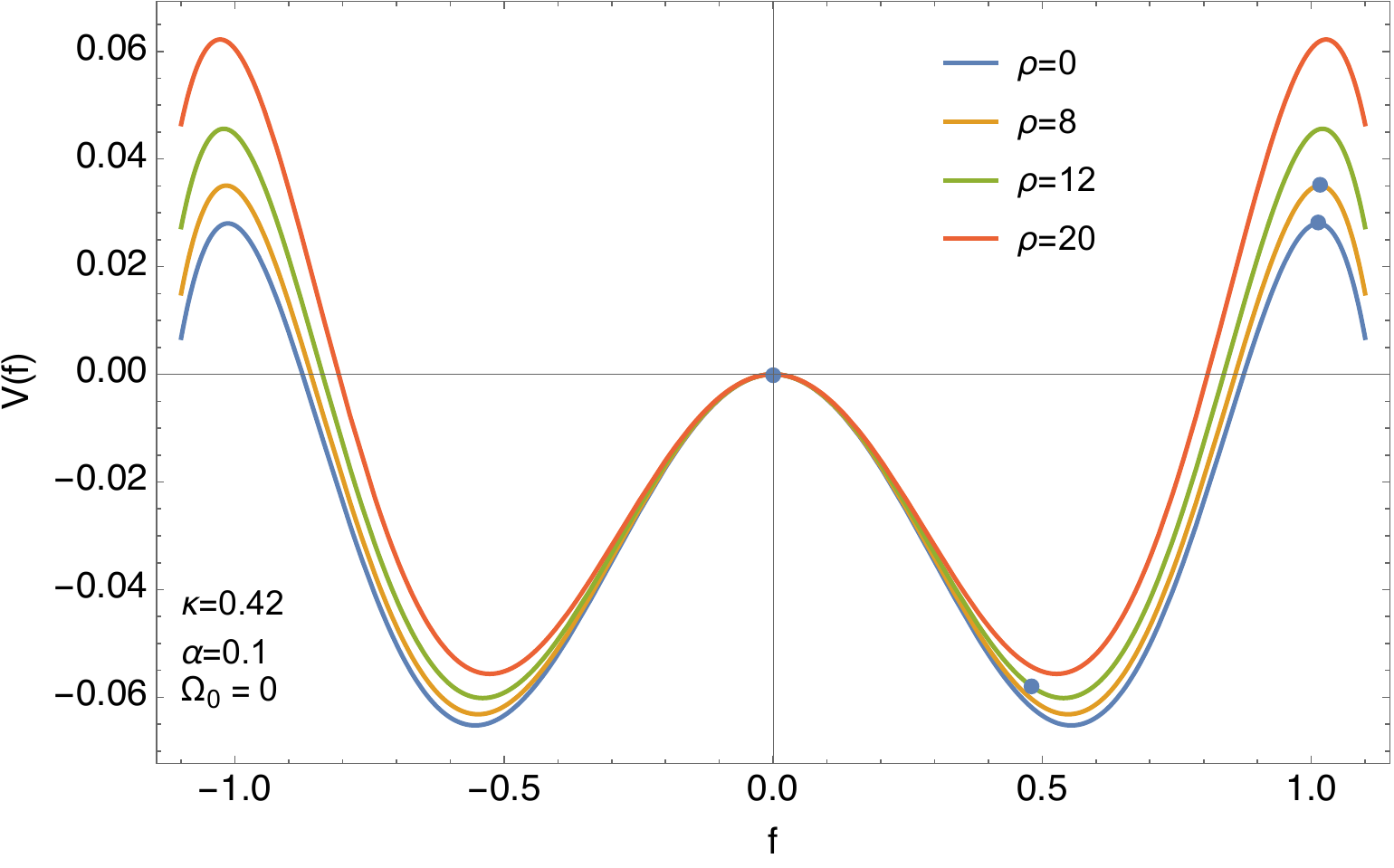}
\includegraphics[width=0.485\textwidth]{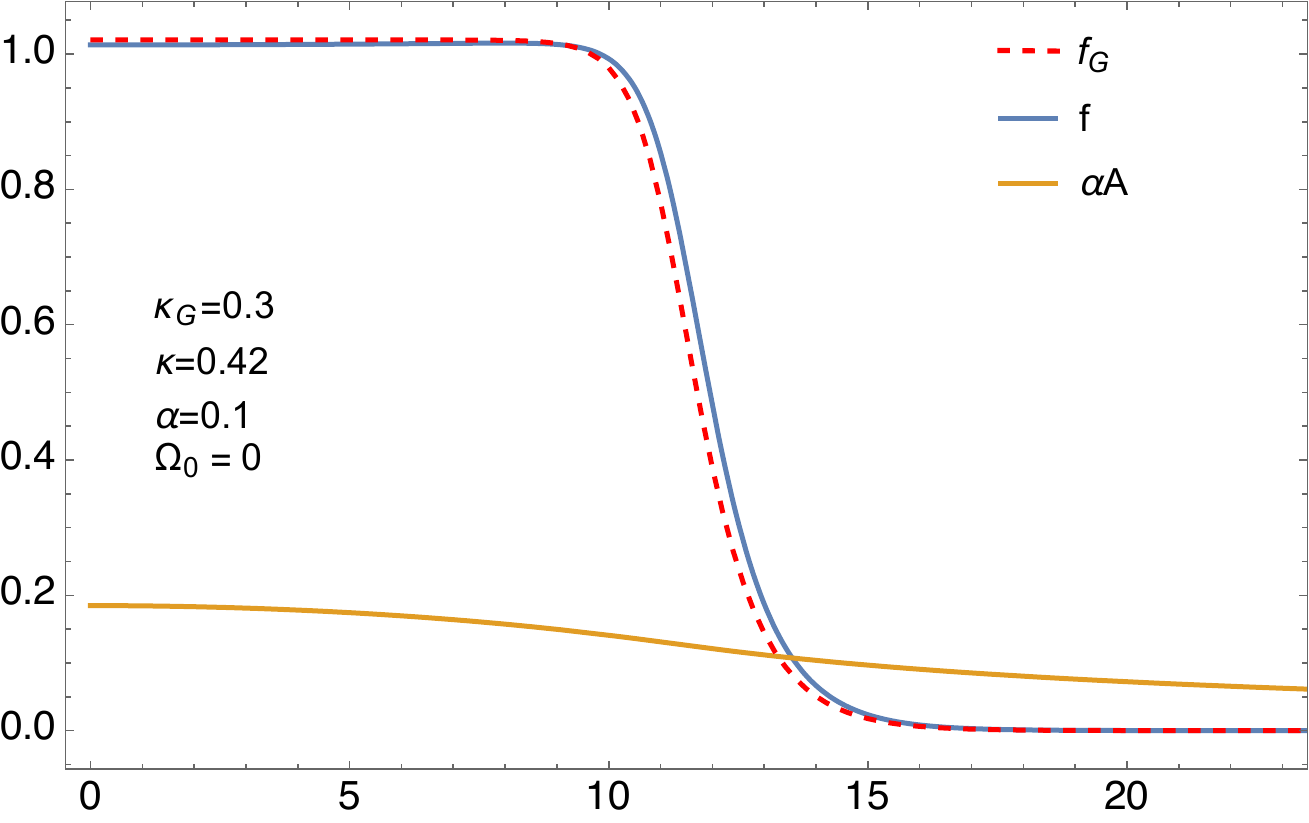}
\caption{Effective potential of the gauged Q-ball changing with $A(\rho)$ vs $f$ and the black dots representing the value of the scalar field for each $\rho$ (left). Exact profiles of the gauged Q-ball denoted by the solid lines and the global Q-ball that maps to the specified gauged Q-ball denoted by the dashed line (right).
 }
\label{fig:2}
\end{figure}

Numerically solving the gauged Q-ball problem via the shooting method is doable but rather tedious and describing the properties of these solution is even harder with two field dependent on one another. Therefore, as shown in~\cite{Heeck_2021map}, a successful method in reducing the gauged Q-ball problem is by realizing there is a mapping equation between gauged and global Q-balls, which simplifies the analysis numerically and analytically. The map is derived by first solving the equation of motion of $A$ Eq.(\ref{24}) in the thin-wall limit where the scalar field can be approximated in terms of a step-function $f(\rho)=1-\Theta(\rho-R^*)$~\cite{Lee:1988ag}
\beq
A(\rho)=\frac{\Omega }{\alpha}\left\{\begin{array}{cc}
\displaystyle 1-\frac{\sinh\left(\alpha\rho\right)}{\cosh\left(\alpha R^\ast \right)\alpha\rho}\,, 
& \rho<R^\ast\,,\\[0.4cm]
\displaystyle\frac{\alpha R^\ast-\tanh\left(\alpha R^\ast \right)}{\alpha\rho}\,, & \rho\geq R^\ast\,.\\
\end{array}\right. \label{32}
\eeq
In this limit, when the radius of the Q-ball is large, the change in $\alpha A$ is expected to be small since $\alpha A'<\Omega/R^*$ meaning the scalar field equation of motion at $\rho \sim R^*$ can be written as
\beq
f'' + \frac{2}{\rho} f'=\frac{1}{\Phi_0^2(m_\phi^2-\omega_0^2)^2}\frac{\dd U}{\dd f}-
\left[\Omega-\alpha A(R^\ast)\right]^2f \,,\label{33}
\eeq
which is identical to the global Q-ball equation of motion (Eq.(\ref{7})) if we identify 
\beq
\Omega_G=\Omega-\alpha A(R^\ast)\,.\label{34}
\eeq
By substituting Eq.(\ref{32}) in Eq.(\ref{34}) we get the mapping relation
\begin{align}
\Omega(R^\ast)= \Omega_G(R^\ast)\,\alpha R^\ast \coth(\alpha R^\ast) \,. \label{35}
\end{align}

This relation was derived in Ref.~\cite{Heeck_2021map}, and the numerical and analytical success of the mapping equation, even beyond the thin-wall limit, in describing the gauged Q-ball ground was demonstrated in that article. Regardless of the value of $\kappa$, the finite element method with the mapping relation could numerically produce the ground state of gauged Q-balls profiles (Fig.~\ref{fig:2}). Also, analytical approximations of the maximum radius, charge, and energy of gauged Q-ball ground state, that match the numerical finding, has been derived in the thin-wall limit via the mapping relation. In the next sections, we extend the mapping relation to excited states of gauged Q-balls in order to expand the numerical and analytical analysis to cover the full gauged Q-ball space.

\section{Excited Gauged Q-balls Profiles}
\label{III}
To numerically compute excited gauged Q-ball profiles using the finite element method~\cite{Heeck_2021map,Panin:2016ooo,Mathematica} in \emph{Mathematica}, we re-write our equations of motion in terms of our compactified coordinate $y$
\begin{align}
&\left(1-\frac{y}{a} \right)^4\left(f''+\frac{2}{y}f' \right)+f\left(\kappa^2+\alpha A(\alpha A-2\Omega)
-1+4f^2-3f^4 \right)=0 \,,\\
&\left(1-\frac{y}{a} \right)^4\left(A''+\frac{2}{y}A' \right)-\alpha f^2\left(\alpha A-\Omega \right)=0\,,
\label{36}
\end{align}
where $y=\frac{\rho}{1+\frac{\rho}{a}}$ and $a$ is a positive number, which is assumed to be large compared to the Q-ball radius. Therefore, the boundary conditions in terms of our new coordinate are
\begin{align}
f(a)=f'(0)=A(a)=A'(0)=0\,.
\label{38}
\end{align}

\begin{figure}[H]
\includegraphics[width=0.48\textwidth]{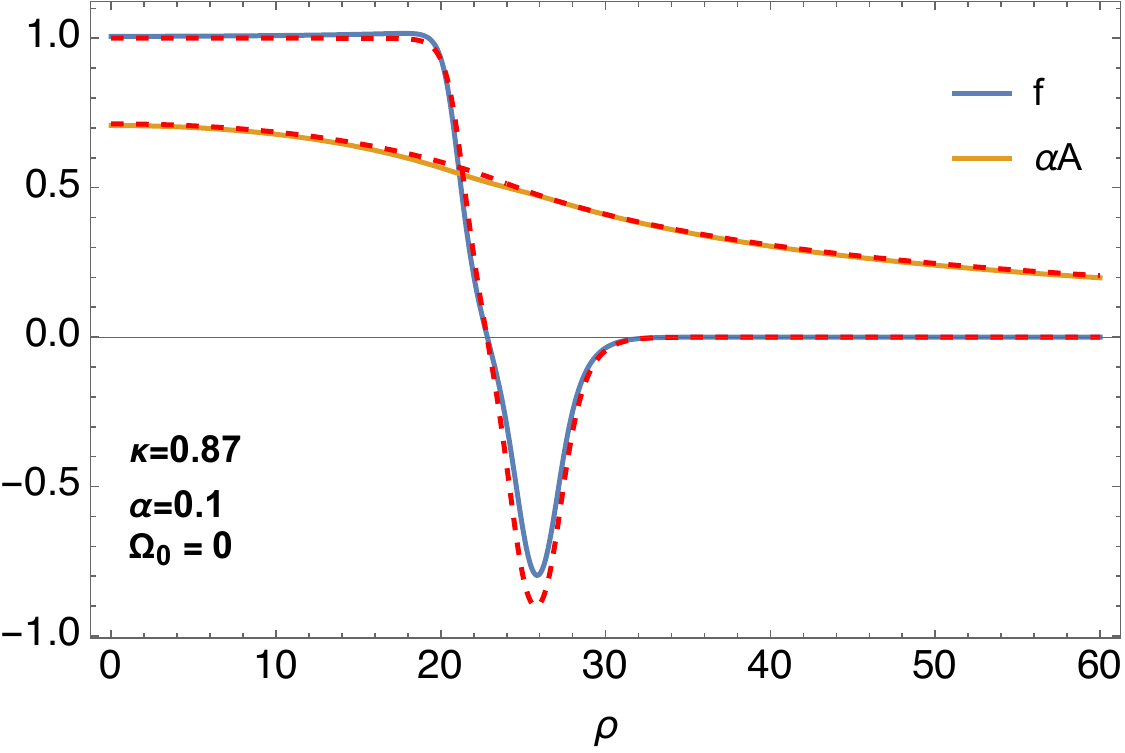}
\includegraphics[width=0.48\textwidth]{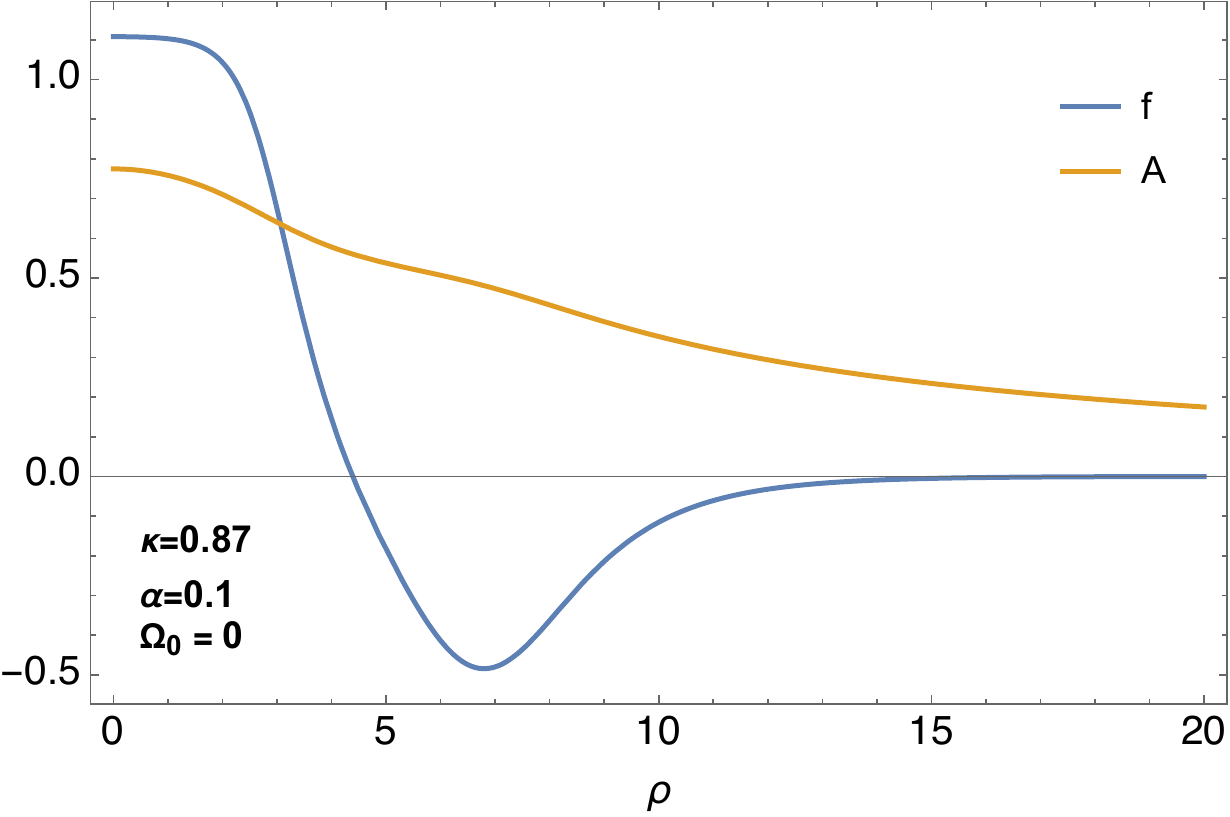}
\caption{Exact profile of first excited state of gauged Q-ball in the thin-wall limit denoted by the solid line, and the thin-wall approximations from Eq.(\ref{14}) and Eq.(\ref{32}) are denoted by the dashed line (left). Exact profiles of first excited state of gauged Q-ball beyond the thin-wall limit denoted by the solid line (right).
 }
\label{fig:3}
\end{figure}

The difference between the the ground state and the excited state is in the initial seed function for the scalar field $f$, and the radius $R^*_N$ relationship to $\kappa_G$ as shown in Eq.(\ref{60}). The mapping equation (Eq.(\ref{35})) is successful in computing $\kappa$ for all possible excited states of gauged Q-balls with coupling $\alpha$, and $\Omega_0$ parameter. Even though excited (global/gauged) Q-balls have $(2N+1)$ radii satisfying $f''(R_{N,n})=0$ as pointed in the review section, $R^*_N$ is enough to determine the properties of these configurations. Numerically, this is illustrated by Fig.~\ref{fig:3} where excited gauged Q-ball profiles are obtained by only specifying $R^*_N$ in the finite element method code and this is generally true for all excited  states. In Fig.{~\ref{fig:3}} we also show that Eq.(\ref{14}) and Eq.(\ref{32}) provide good approximations of the scalar $f$ and gauge $A$, in the thin-wall limit, and the approximations break beyond that limit. To use Eq.(\ref{14}) as shown in Fig.{~\ref{fig:3}}, one needs to provide $R_{N,n}$, which can either be extracted from  the numerical profile at $f''(R_{N,n})=0$ or analytically predicted using Eq.(57) from Ref.\cite{Almumin_2022}. Numerical solutions do not exist for arbitrary large $N$ excitations of gauged Q-balls, as illustrated in Ref.~\cite{Loginov_2020}, and in the next section we will use the mapping relation to approximate an analytical upper bound on the maximum number of possible excited states per gauged Q-ball. 

\section{Approximations of Excited Gauged Q-balls Properties}
\label{IV}
The mapping relation cannot be solved analytically, however, approximations have been deduced for unexcited gauged Q-balls by understanding the limiting cases of the mapping function~\cite{Heeck_2021map}. And as we showed in the previous section, the mapping equation can be extended to predict excited states, therefore, we try to find generalized analytical approximations of our gauged Q-ball space. 
Recall that in the thin-limit, the radius (characterizing length scale) of excited Q-balls is approximated by $R^*_N=(2N+1)/\kappa_G^2$ (Eq.(\ref{60})), thus, the map can be re-written in terms of $\kappa_G$  
\begin{equation}
   \kappa^2=\frac{\tilde{\alpha}^2}{\kappa_G^2} (1+\frac{\Omega_0^2}{\kappa_G^2}) \text{coth}^2(\frac{\tilde{\alpha}}{\kappa_G^2} ) -\Omega_0^2\,,
   \label{39}
\end{equation}
where $\tilde{\alpha}=(2N+1)\alpha$. Notice that the equation is now identical to the ground state mapping relation. This allows us to use the analytical expression that was derived in~\cite{Heeck_2021map}
\begin{align}
    \tilde{\alpha} \lesssim \frac{1}{\sqrt{1/0.58^2+9\Omega_0^2/2}}\,,
    \label{40}
\end{align}
in order to further understand the excited gauged Q-ball properties. By re-writing the the bound in terms of $\alpha$ and $N$ 
\begin{align}
    N_{\text{max}} \lesssim \frac{1}{2\alpha \sqrt{1/0.58^2+9\Omega_0^2/2}}-\frac{1}{2}\,,
    \label{41}
\end{align}
we obtain an upper limit on the number of the excited states a gauged Q-ball could acquire. This bound agrees with the findings from Ref.~\cite{Loginov_2020} where it was shown that excited gauged Q-balls have a finite number of excited states, which is inversely proportional to the coupling implying that a minimum coupling value ($\alpha\approx 0.182$) is required for an excited state to exist. Now, we have an analytical expression which can quantitatively approximate the maximum number of excited states a gauged Q-ball can acquire depending on the  gauged coupling $\alpha$ and the parameter $\Omega_0$.

The mapping equation clearly implies that $\kappa$ is larger for higher excited states of gauged Q-balls with a certain radius since $\tilde{\alpha}$ increases with $N$. Also, the nature of coth function Eq.(\ref{39}) coupled with the bound $\kappa \leq 1$ entail a decrease in the radii space for higher excitation. By imposing the upper bound ($\kappa=1$), we get the following equation
\begin{equation}
   1=\frac{\tilde{\alpha}^2}{\kappa_G^2} \left(1+\frac{\Omega_0^2}{\kappa_G^2}\right) \text{{coth}}^2\left(\frac{\tilde{\alpha}}{\kappa_G^2} \right) -\Omega_0^2\,,
   \label{42}
\end{equation}
  which allows us to analytically approximate the maximum $R^*_{N({\text{max}})}$ (thin-wall limit) in terms of $\alpha$, $\Omega_0$ and $N$. As mentioned earlier the mapping equation is not analytically solvable but we can deduce $R^*_{N({\text{max}})}$ limits for regions when $\alpha \gtrsim \Omega_0$ and $\tilde{\alpha}<<\Omega_0$. In the first case, the maximum radius is approximated to be
\begin{align}
   R^*_{N({\text{max}})}\simeq\frac{1}{\left(2N+1\right)\alpha^2}\,, \ \ \ \ \ \alpha\gtrsim \Omega_0\,,
    \label{43}
\end{align}
for all $N$ excited states with $\alpha$ gauged coupling. This analytical approximation works because when substituting in the mapping equation it reduces to
\begin{align}
  \left(1+\frac{\Omega_0^2}{(2N+1)\alpha^2}\right) \text{coth}^2\left(\frac{1}{(2N+1)\alpha}\right)-\Omega_0^2=1\,,
  \label{44}
\end{align}
where the second and third term are small compared to the first term and the second term gets smaller for larger $N$. The approximation works for arbitrary $\alpha$ and $N$ since the number of excited states are inversely proportional to the gauge  coupling as demonstrated by Eq.(\ref{41}). This relation implies that $\text{coth}^2(1/((2N+1)\alpha)) \approx 1$ for all $N$ as there will not be enough excited states per gauged coupling that will increase the value of the function significantly. For the second case when $\tilde{\alpha}<<\Omega_0$ ground state limit for the maximum
\begin{equation}
   R^*_{N({\text{max}})}\simeq\frac{1}{\alpha \Omega_0}\,, \ \ \ \ \ \tilde{\alpha}<<\Omega_0\,,
    \label{45}
\end{equation}
holds. In the second case, unlike the first one, the approximation does not work for arbitrary $N$ but is restricted by $(2N+1){\alpha}<<\Omega_0$ and this can be seen by substituting the radius again in the mapping equation (Eq.(\ref{35}))
\begin{equation}
    \left(\frac{\alpha(2N+1)}{\Omega_0}+1\right)\text{coth}^2\left(\frac{1}{\Omega_0}\right)-\Omega^2_0=1\,,
    \label{46}
\end{equation}
The second  term is independent of $N$ and clearly larger than the first term for the specified region allowing for the approximation of the maximum radius to hold.

 \begin{figure}[H]
 \centering
\includegraphics[width=0.485\textwidth]{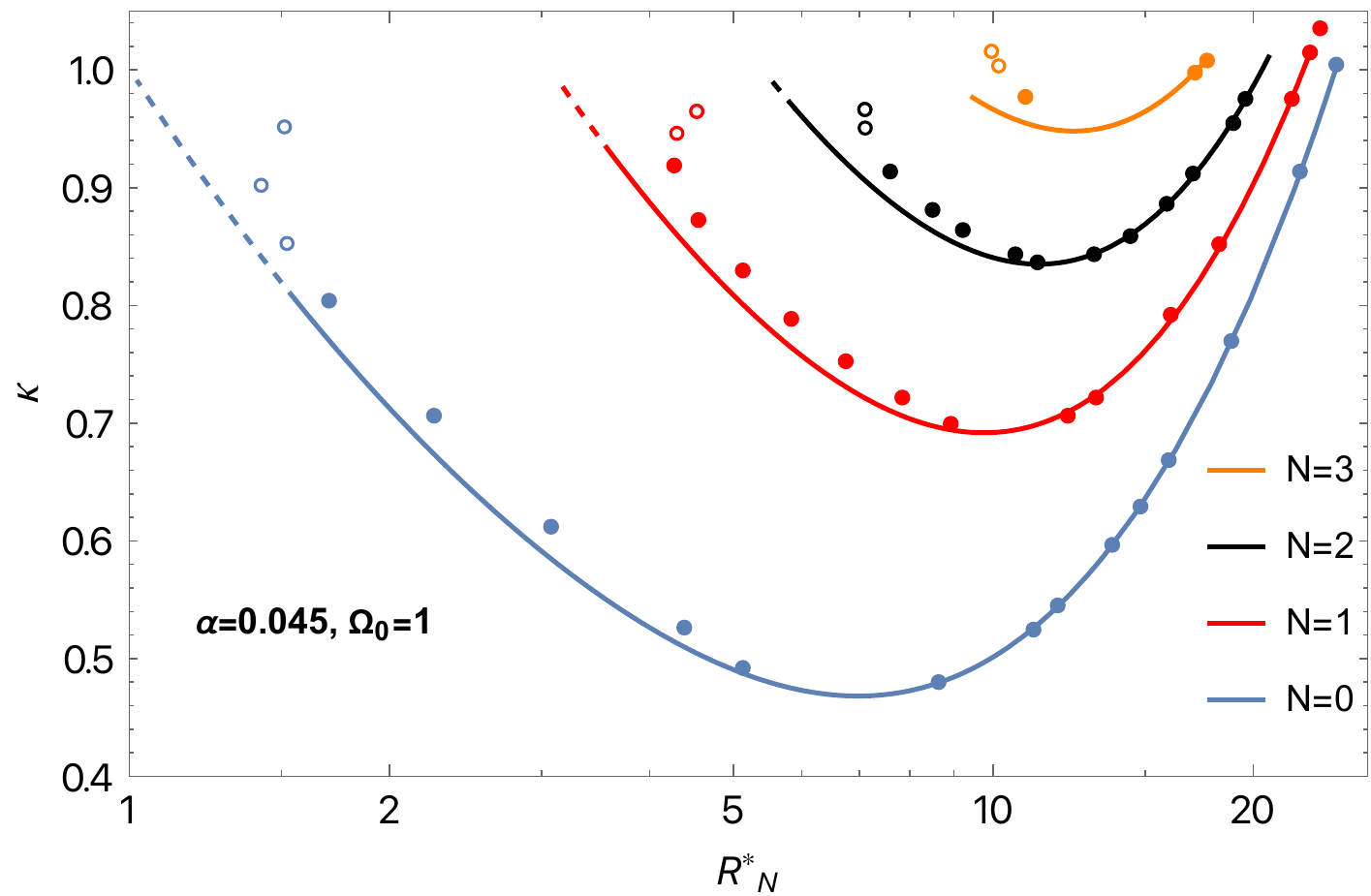}
\includegraphics[width=0.485\textwidth]{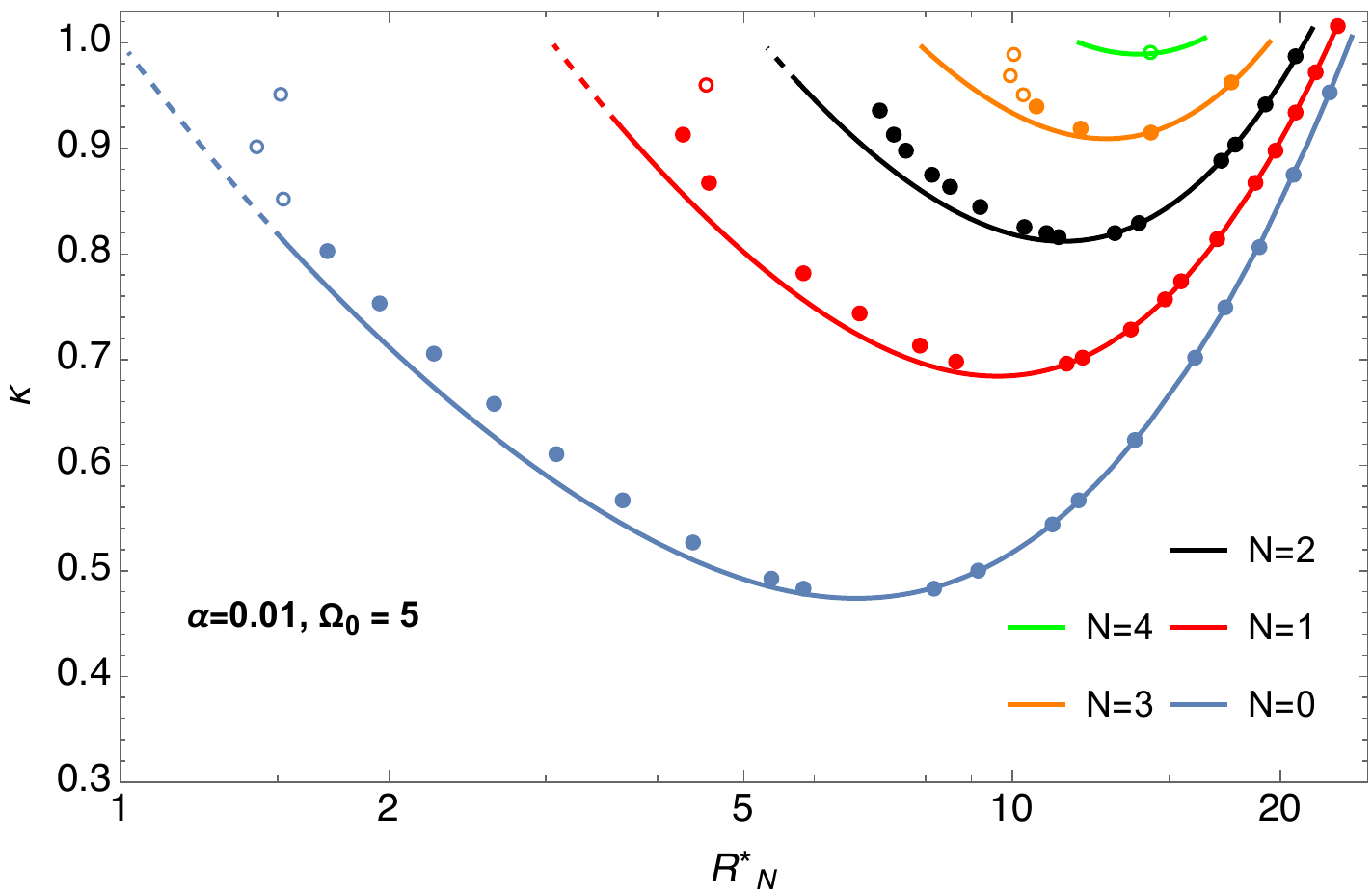}
\includegraphics[width=0.5\textwidth]{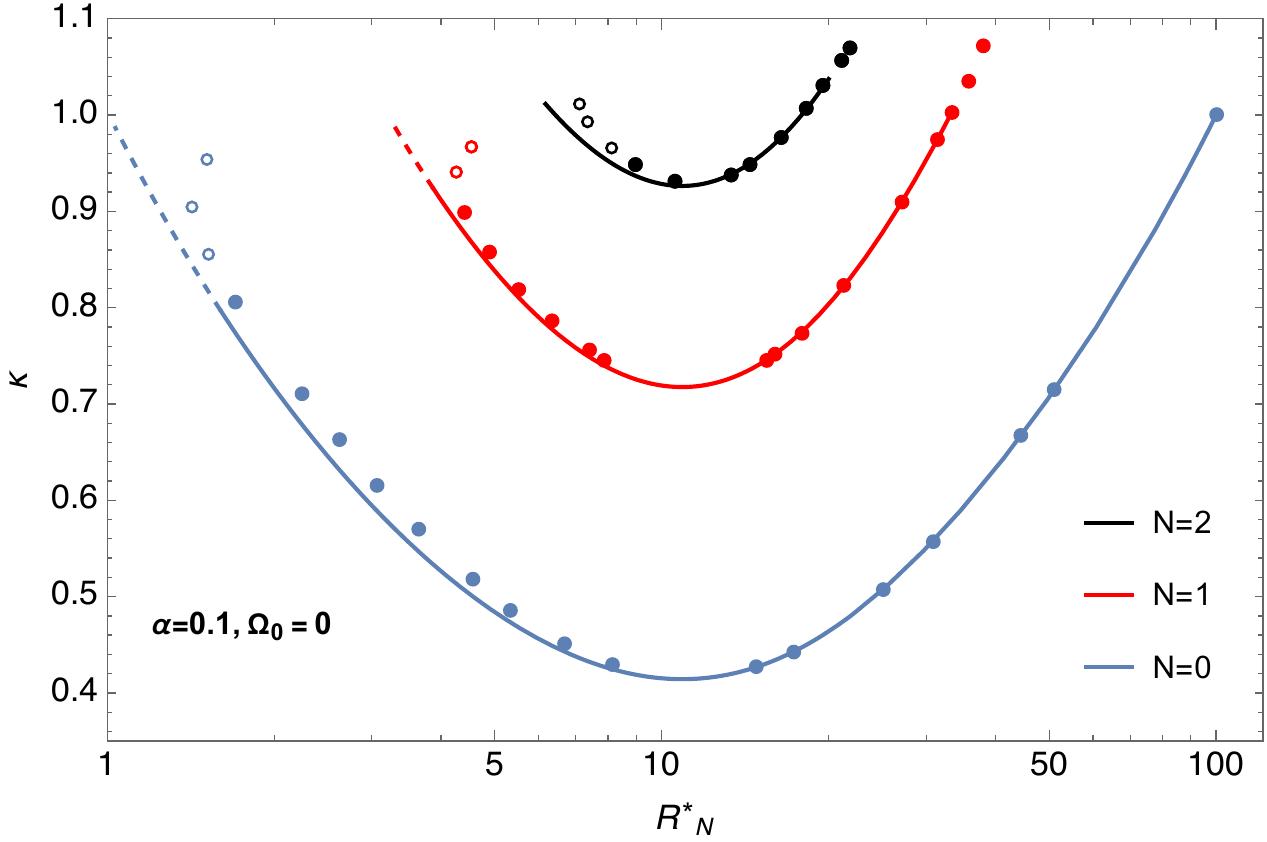}
\caption{Exact values of $\kappa$ vs $R^*_N$ for all possible excited states of gauged Q-balls denoted by the dots compared to the analytical prediction from the mapping equation and the radii approximation from Eq.(\ref{39}) and Eq.(\ref{60}). Different benchmarks are shown: $\alpha=0.045$, $\Omega_0=1$ (top left), $\alpha=0.01$, $\Omega_0=5$ (top right), and $\alpha=0.1$, $\Omega_0=0$ (bottom). Solid lines and dots denotes the Q-ball region where $E \leq m_\phi Q$ is satisfied.  
}
\label{fig:4}
\end{figure}

 We chose three different combinations of $\alpha$ and $\Omega_0$ in Fig.~\ref{fig:4} to demonstrate the approximated limits we found for maximum number of excited states a gauged Q-ball could acquire and the maximum radius admitted per state. For each state analytical $\kappa$ vs $R^*_N$ has been plotted using Eq.(\ref{60}) for the radii and the mapping relation Eq.(\ref{36}) for $\kappa$. Smaller radii are expected to have higher order corrections that are harder to obtain, however, the thin-wall approximation is sufficient for our discussion. Exact numerical radii of excited gauged Q-balls are represented by the dots in the figure and they match very well with the analytical solution. Finding distinct exact solutions becomes harder in the region where the variation in the value of $\kappa$ is very small (Fig.~\ref{fig:4}). 
 
 The upper limit on $N_\text{max}$ is illustrated numerically and analytically for different $\alpha$ and $\Omega_0$ benchmarks in Fig.~\ref{fig:4}. For $\alpha=0.01$ and $\Omega_0=5$ benchmark, the maximum possible excited state would be at $N_\text{max}=4$ as shown analytically and numerically, which agrees with Eq.(\ref{41}) where $N_\text{max}\lesssim 4.15$. Also, Fig.~\ref{fig:4} demonstrates our finding from Eq.(\ref{43}) and Eq.(\ref{45}). In the the example where $\Omega_0=0$ the maximum radius decreases with larger excited states and matches our approximation from Eq.(\ref{43}). However, when $\Omega_0\gg\tilde{\alpha}$ it is apparent that the maximum radii cluster around the limit we show in Eq.(\ref{45}). As mentioned earlier, the second approximation does not hold for all $N$, and this is apparent in $N=3$ excited state when $\Omega_0=1$. 
 
It is evident from our discussion that the space of possible solutions per excited gauged Q-ball shrinks for higher states. We can infer qualitatively from the mapping equation and $\kappa \leq 1$ that the minimum radius increases for higher excited states, even though finding an analytical expression for the minimum radius is complicated beyond the thin-wall limit. The smaller the region of possible radii and the $\kappa$, the harder it is to find distinct numerical solutions, which explains the small number of exact solutions provided for higher excited states in Fig.~\ref{fig:4}. We re-iterate that the radius approximation we use does not capture the exact behaviour for smaller radii since we expect higher order terms to become relevant in this region for excited states adjusting the shape of the radii curves just like the  ground states shown in Ref.~\cite{Heeck_2021map}. This is going to be beyond the scope of this paper, but could be an interesting topic to investigate to improve the findings beyond the thin wall limit. In the plots you would notice that the curves and dots are divided into solid and dashed segments representing different instability regions, which we will discuss in the next section.

\section{Charge, Energy and Stability}
\label{V}

Excited gauged Q-balls are unstable configurations, and to determine the type of instability and where a occurs, we must discuss the charge and energy of of these states. The charge $Q$ and energy $E$ of gauged Q-balls as shown in Sec.\ref{II} are given by 
\begin{equation}
    Q= \frac{4 \pi \phi_0^2}{m_\phi^2-\omega^2_0}\int d\rho\rho^2 f^2 (\Omega-\alpha A),
    \label{47}
\end{equation}
\begin{equation}
    E=\omega Q+\frac{4\pi \phi_0^2}{3\sqrt{m_\phi^2-\omega_0^2}}\int d\rho \rho^2 (f'^2-A'^2).
    \label{48}
\end{equation}

\begin{figure}[t]
\includegraphics[width=0.49\textwidth]{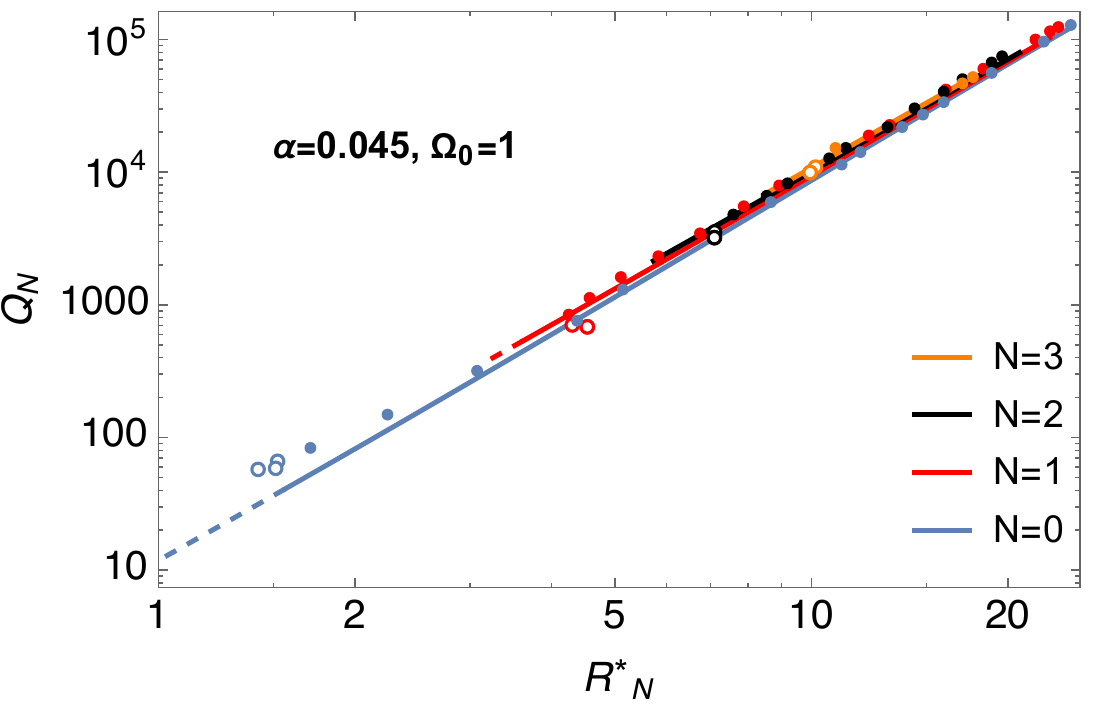}
\includegraphics[width=0.49\textwidth]{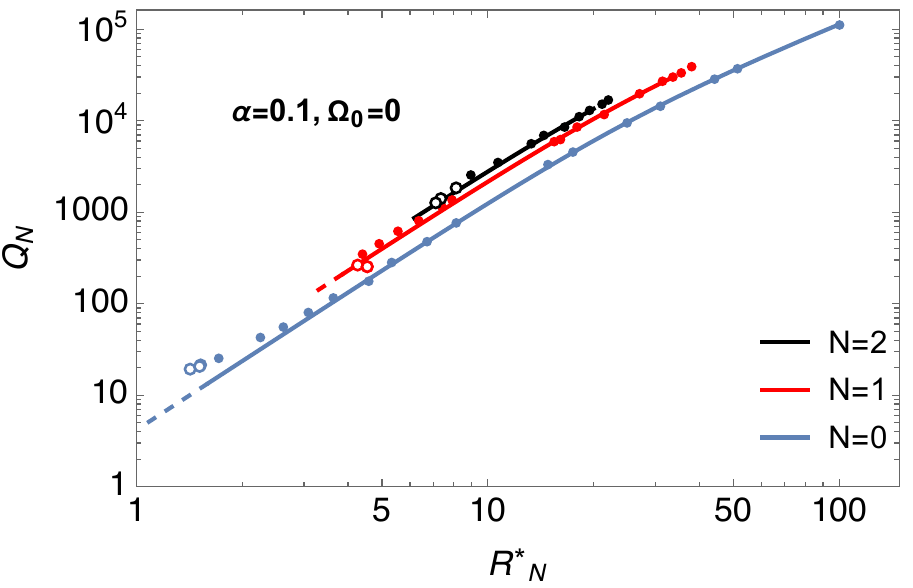}
\includegraphics[width=0.49\textwidth]{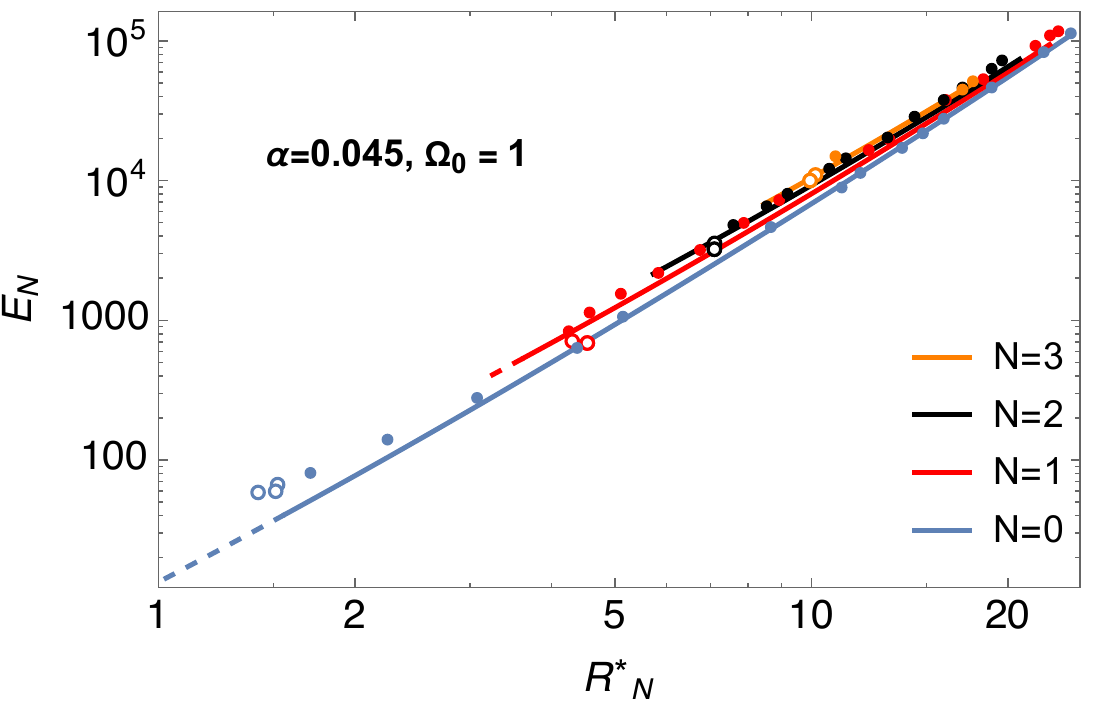}
\includegraphics[width=0.49\textwidth]{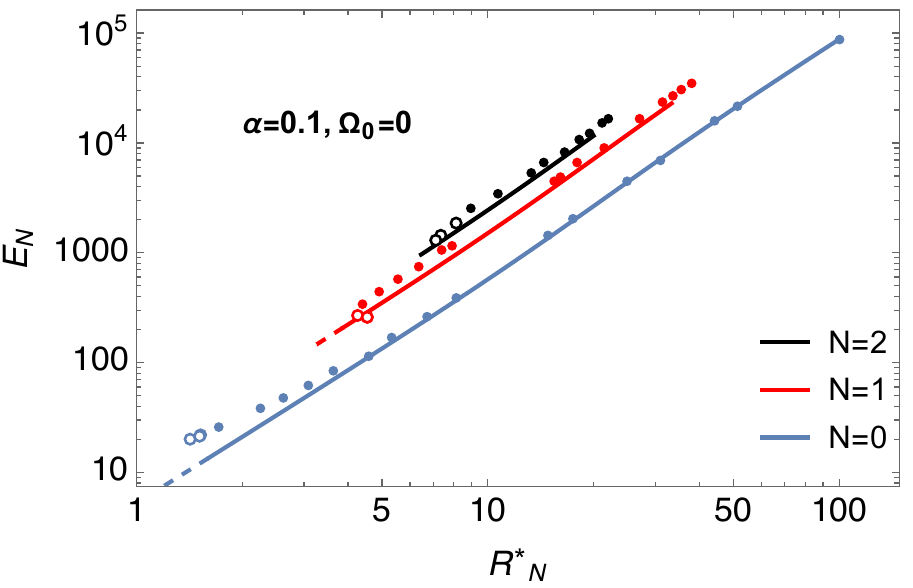}
\includegraphics[width=0.49\textwidth]{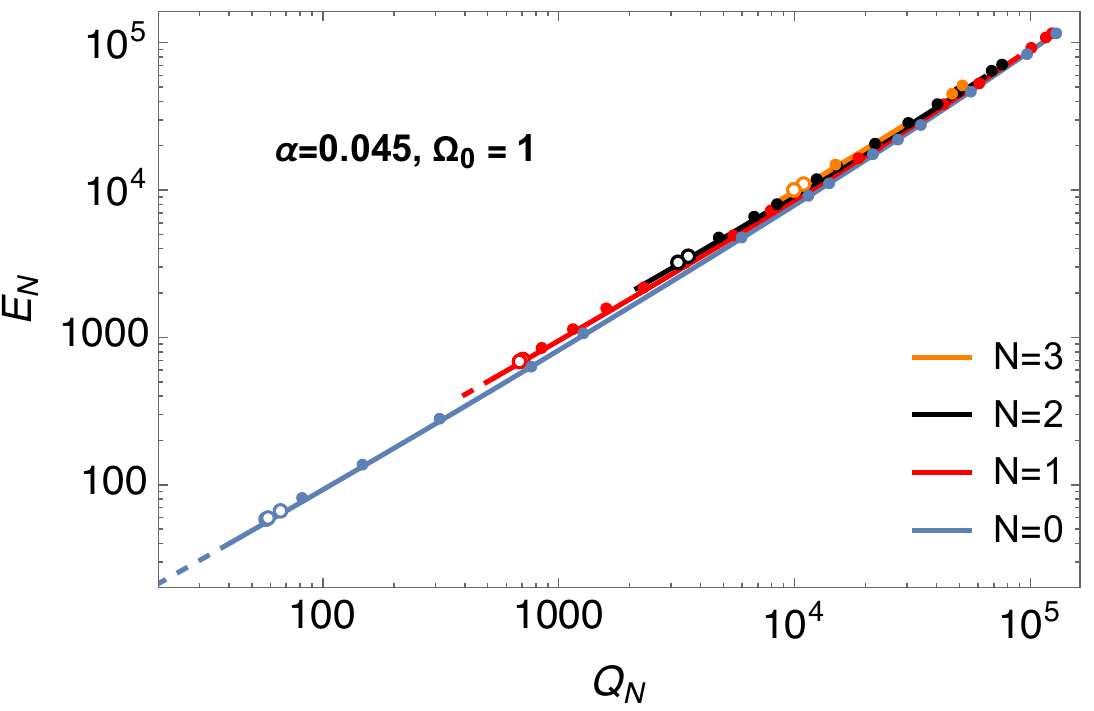}
\includegraphics[width=0.49\textwidth]{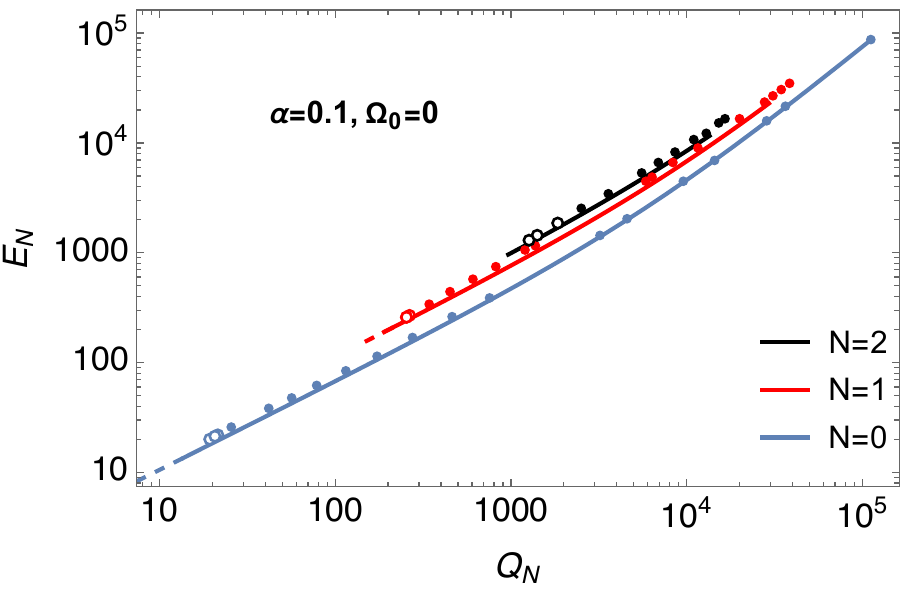}
\caption{Exact values of charge and energy vs $R^*_N$ and energy vs charge for all possible states of gauged Q-balls denoted by the dots compared to the analytical thin-wall predictions from  Eq.(\ref{49}-\ref{50}). Different benchmarks are shown: $\alpha=0.045$, $\Omega_0=1$ (left), and $\alpha=0.1$, $\Omega_0=0$ (right). Solid lines and dots denotes the Q-ball region where $E \leq m_\phi Q$ is satisfied and $m_\phi=\phi_0=1$ for all benchmarks.  
 }
\label{fig:5}
\end{figure}

The exact integrals are numerically calculable over the profiles we produced in Sec.~\ref{III} and shown as dots in Fig.~\ref{fig:5}. Since the mapping relation worked for excited states, the approximation of charge and energy derived in~\cite{Heeck_2021map} of the ground state of gauged Q-ball in the large radius limit could be extended to estimate the observables for the excited states
\begin{equation}
    Q_N=\frac{4\pi \Omega \Phi_0^2}{\alpha^3}\left(\alpha R_N^*-\text{tanh}\,(\alpha R_N^*)\right)
    \label{49}
\end{equation}
\begin{equation}
    E_N=\omega Q_N +\frac{\pi \phi_0^2 R_N^{*2}}{3\sqrt{m_\phi^2-\omega_0^2}}-\frac{\pi \phi_0^2}{3\sqrt{m_\phi^2-\omega_0^2}}\frac{\Omega^2(\alpha R_N^* (\text{sech}(\alpha R_N^*)+2)-3\text{tanh}(\alpha R_N^*))}{2\alpha^3}\,,
    \label{50}
\end{equation}

 as shown in Fig.~\ref{fig:5}. This demonstrates the point we made in Sec.{~\ref{III}} where we argued that properties of excited gauged Q-balls are determined by the value of $R^*_N$ despite the states having $(2N+1)$ radii satisfying $f''(R_{n,N})=0$.

As we have seen in the previous section, $\kappa$ of gauged Q-balls increase for higher excited states of the same radius, which implies that the charge and energy increase for higher excited states. Curiously though, larger $\Omega_0$ will dominate the expressions $\Omega^2=\kappa^2+\Omega_0^2$ decreasing the influence of changing $\kappa$ for higher excited states. This is shown in Fig.\ref{fig:5}, where we see that the charge and energy for the example with $\Omega_0=0$ increase for higher excited states. However, when increasing $\Omega_0=1$, the distinction between the charge and energy of each state diminishes. This implies that excited states of gauged Q-balls with larger $\Omega_0$ would have have a longer lifetime since the energy gap $\Delta E$ gets smaller with the ground state and the lifetime is inversely proportional to the energy gap. This is illustrated by the $E$ vs $Q$ plot in Fig.\ref{fig:5} this is because when comparing energies of different level of excitations of the \emph{same} gauged Q-ball, we should describe the gauged Q-ball with the \emph{same} charge. It is important to emphasize that these approximations are better for larger $R^*_N$ or smaller $\kappa$, however, they deviate from the exact values beyond the limit.

\begin{figure}[H]
\includegraphics[width=0.49\textwidth]{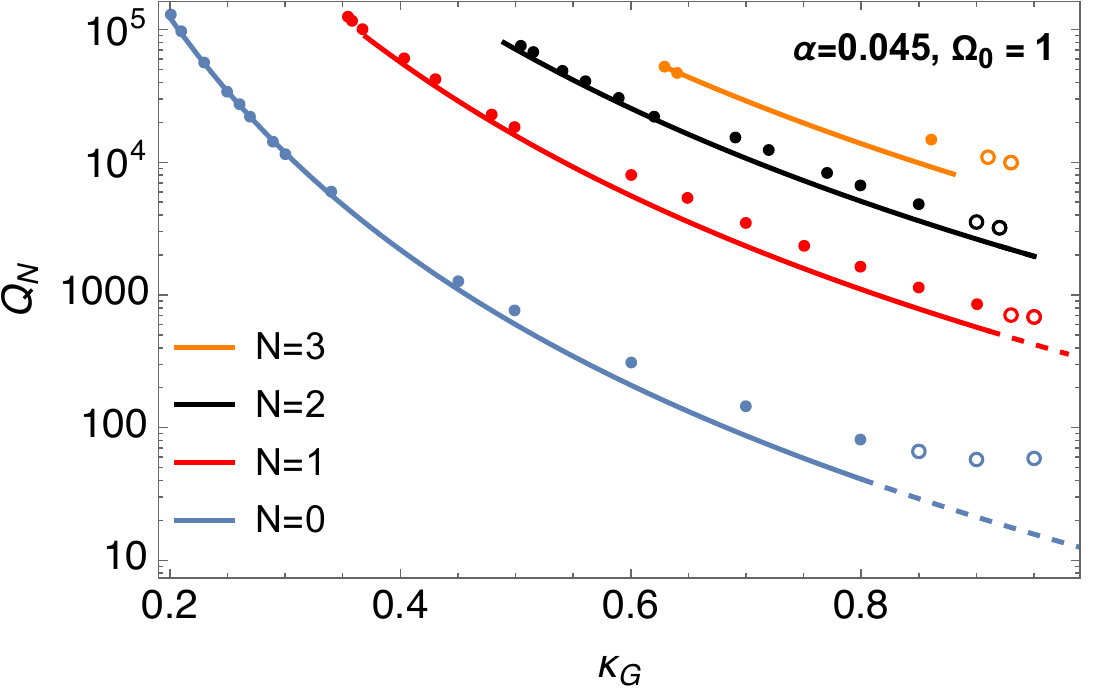}
\includegraphics[width=0.495\textwidth]{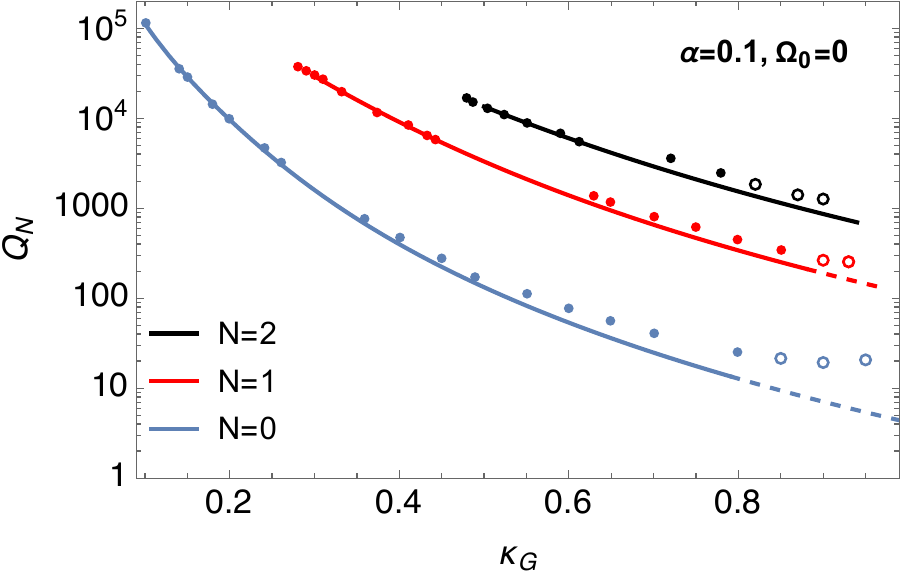}
\includegraphics[width=0.49\textwidth]{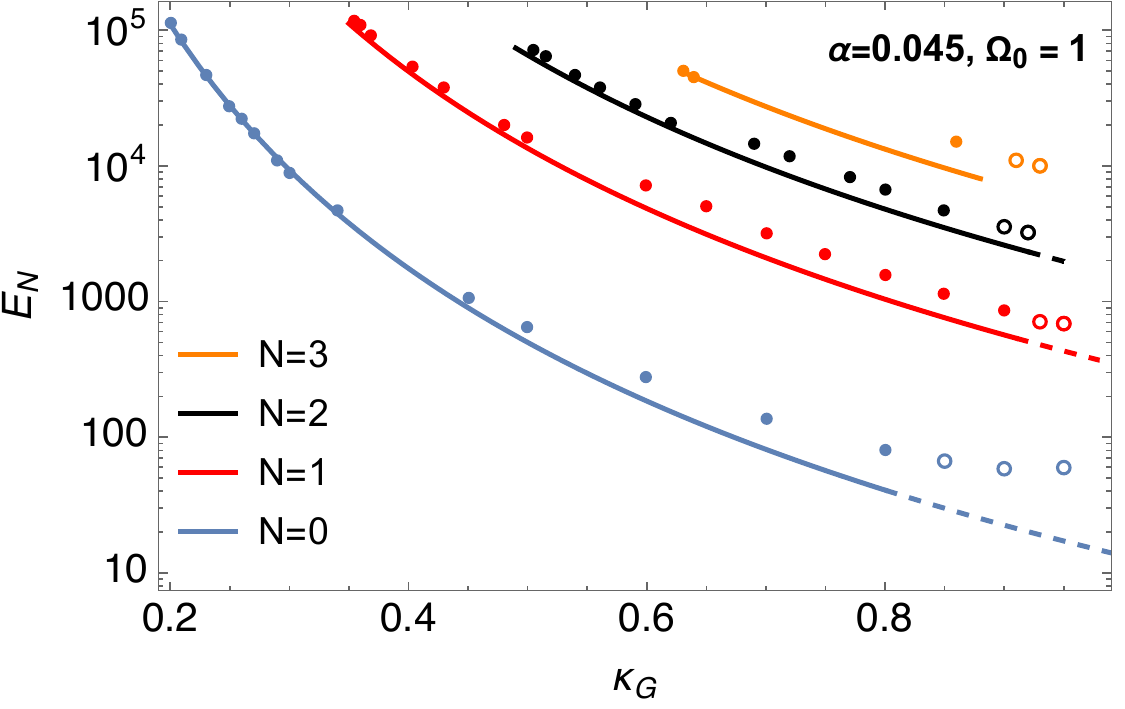}
\includegraphics[width=0.49\textwidth]{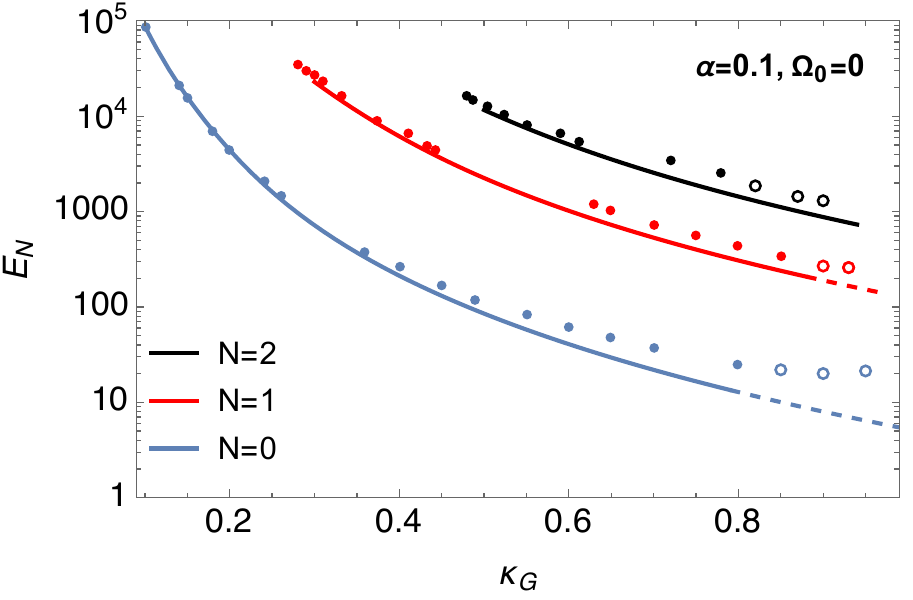}
\caption{Exact values of charge and energy vs $\kappa_G$ for all possible excited states of gauged Q-balls denoted by the dots compared to the analytical thin-wall predictions from  Eq.(\ref{51}-\ref{52}). Different benchmarks are shown: $\alpha=0.045$, $\Omega_0=1$ (left), and $\alpha=0.1$, $\Omega_0=0$ (right). Solid lines and dots denotes the Q-ball region where $E \leq m_\phi Q$ is satisfied and $m_\phi=\phi_0=1$ for all benchmarks. 
 }
\label{fig:6}
\end{figure}

Another way of discussing the charge and energy of excited Q-balls would be in terms of the number of excitations, which is going to be useful to discuss stability. If we think of our charge and energies integrals in terms of $\kappa_G$, a natural relation emerges between the states. The gauged Q-ball charge $Q$ (Eq.(\ref{47}) can essentially be written in terms of the global Q-ball charge $Q_G$ since the map Eq.(\ref{34}) tells us that $\Omega_G=\Omega-\alpha A$. The difference between the charge of the gauged Q-ball and the global Q-ball is that in the gauged case the charge has an upper bound due to the existence of a maximum radius as demonstrated in the previous section. The integral approximation of $\rho^2 f^2$ shown in Eq.(\ref{17}) implies that the charge of excited gauged Q-balls are expressed as 
\begin{equation}
    Q_N(\kappa_G)=(2N+1)^3 Q_0(\kappa_G) \propto \frac{(2N+1)^3}{3\kappa^6}\,,
    \label{51}
\end{equation}
where $Q_0(\kappa_G)$ is the charge of the ground state. Similarly, the energy of excited gauged Q-balls can be written in terms of the ground state energy $E_0(\kappa_G)$ by using the integral approximation of $\rho^2 f'^2$ from Eq.(\ref{18})
\begin{equation}
    E_N(\kappa_G)=(2N+1)^3 E_0(\kappa_G) \propto \frac{(2N+1)^3}{4\kappa_G^4}\,,
    \label{52}
\end{equation}
This approximation evidently ignores the contribution from $A'$ integral from Eq.(\ref{50}) and is valid only when $\kappa_G\gtrsim0.2$, otherwise the the approximation breaks down as the the contribution becomes significant. This approximation could still be relevant for excited gauged Q-balls despite the lower bound on $\kappa_G$ since the maximum radius is expected to get smaller for higher excite states, as shown in Eq.(\ref{43}) and Eq.(\ref{45}) meaning the minimum $\kappa_G$ gets pushed to a larger value. This results captures the significance of the mapping relation where properties that are harder to estimate in the gauged Q-ball set-up could be expressed in terms of global Q-ball properties (Fig.~\ref{fig:6}). 

\begin{figure}[H]
\includegraphics[width=0.49\textwidth]{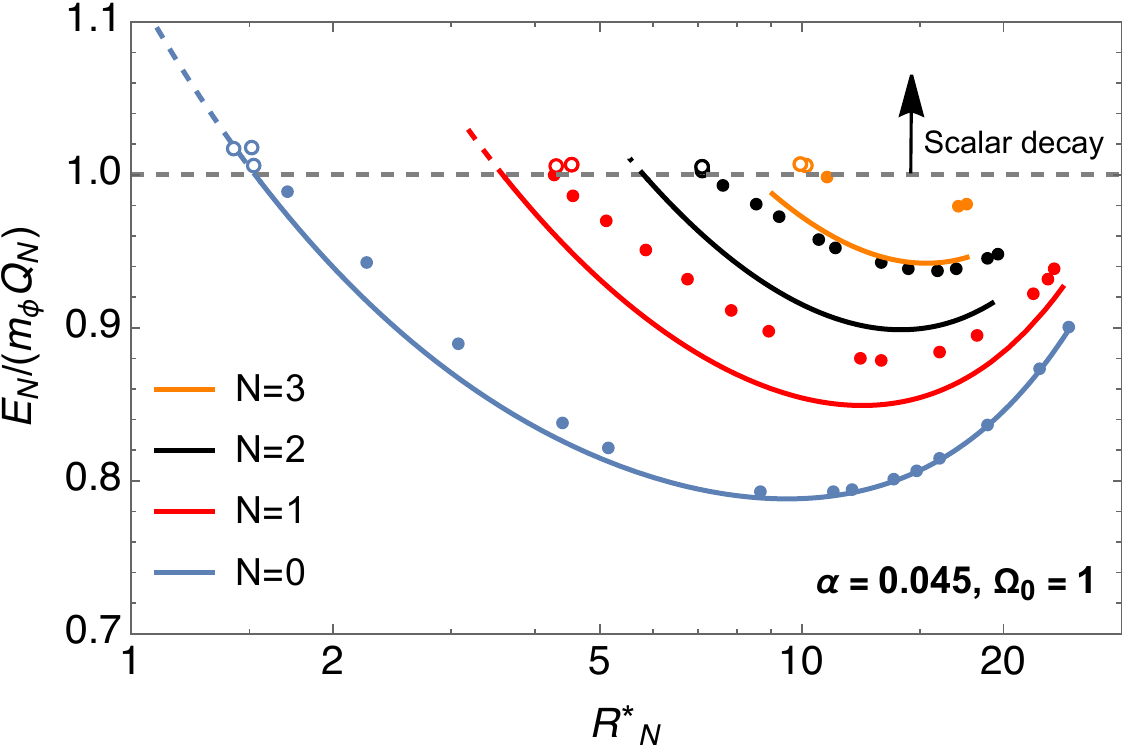}
\includegraphics[width=0.495\textwidth]{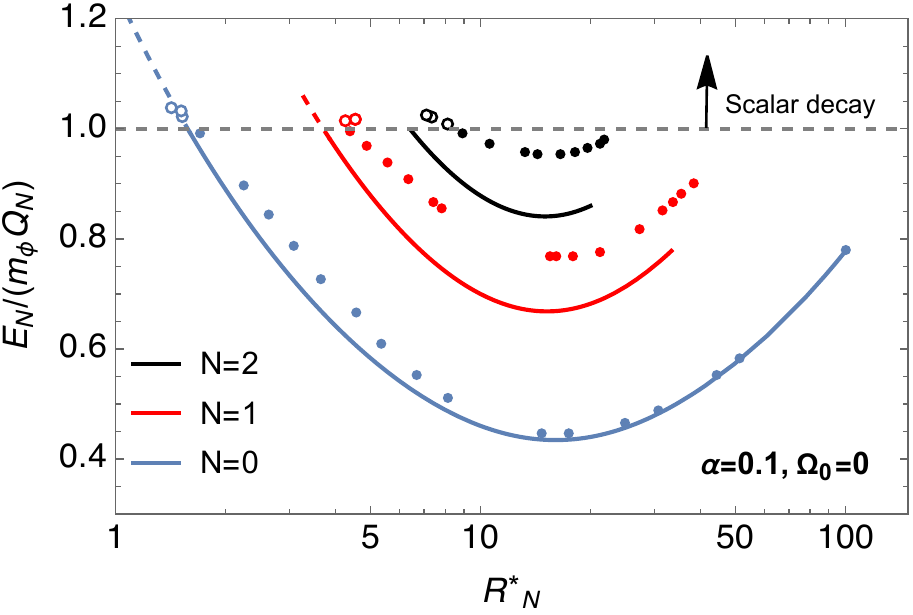}
\includegraphics[width=0.49\textwidth]{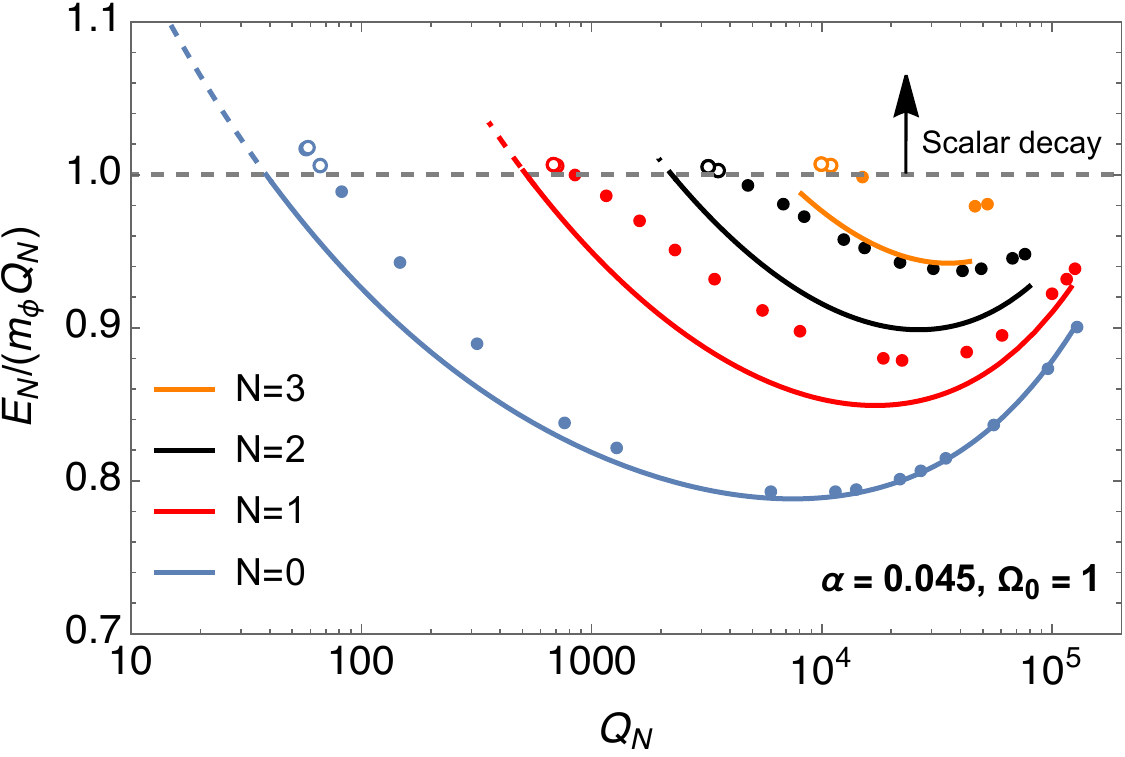}
\includegraphics[width=0.49\textwidth]{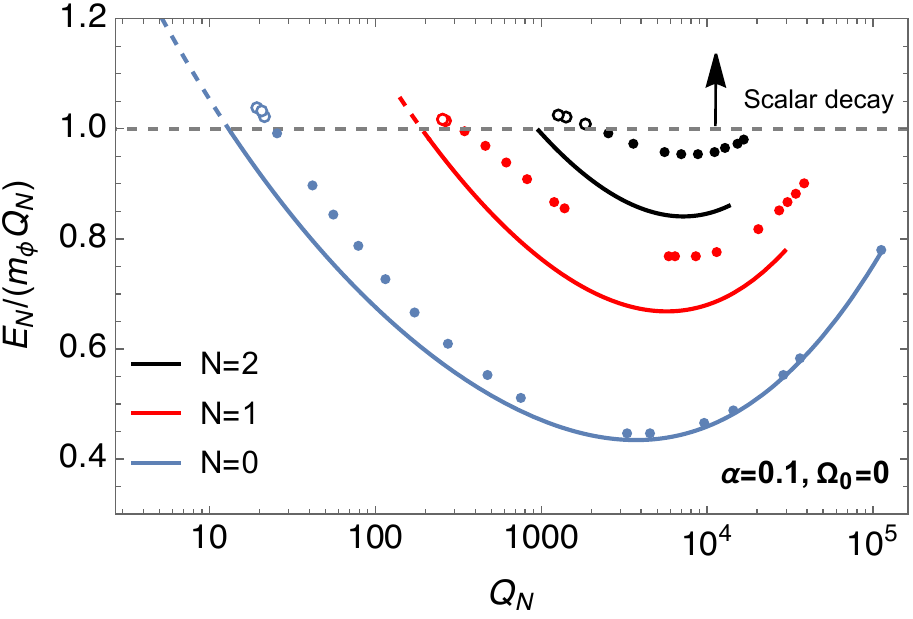}
\caption{Exact values of $E/(m_\phi Q)$ vs $R_N^*$ (top) and $Q$ (bottom) for all possible excited states of gauged Q-balls denoted by the dots compared to the analytical thin-wall approximations from Eq.(\ref{49}-\ref{50}). Excited gauged Q-balls with $E/(m_\phi Q)>1$ decay into Q free scalar, while $E/(m_\phi Q) \leq 1$ decay into ground state Q-balls by emitting free scalars. Different benchmarks are shown: $\alpha=0.045$, $\Omega_0=1$ (left), and $\alpha=0.1$, $\Omega_0=0$ (right). $m_\phi=\phi_0=1$ for all benchmarks. 
 }
\label{fig:7}
\end{figure}

Excited (global and gauged) Q-balls are unstable configurations~\cite{Friedberg:1976me,Heeck_2021map,Loginov_2020}. The first instability comes from the the condition that requires $E\leq m_\phi Q$ for Q-balls to be stable against decay completely into free scalars. The condition is satisfied for the gauged and global Q-ball ground state when $\kappa_G\lesssim0.84$, which implies that the minimal stable radius is expected around $R^*\simeq 1.5$. Therefore, for excited gauged Q-balls we would expect the minimal radius satisfying  $E\leq m_\phi Q$ condition to approximately be around $R^*\simeq (2N+1)/\kappa_{G,\textit{stability}}^2$ if the radius relation to $\kappa$ holds where $\kappa_{G,\textit{stability}}$ increases for excited  states~\cite{Almumin_2022}. However, as we pointed out earlier, in this region higher order corrections become relevant meaning this approximation is not reliable and we should rely on the numerical results. This explains the disagreement in the minimum stable radius between analytical approximation curves and the exact solutions shown in Fig.~\ref{fig:7}, where the exact minimum radius is larger than our approximation.

Even though the Q-ball configuration is stable when $E\leq m_\phi Q$, excited gauged Q-balls are not stable in this region against decay into Q-balls with lower energy by emitting scalars. This fact can be illustrated by writing the energy of the excited gauged Q-balls in term of their charge. Since approximations in Eq.(\ref{51}-\ref{52}) hold in the thin-wall limit, the gauged Q-ball case, similar to the excited global Q-ball case, we could re-write the energy in terms of the charge when $\kappa_G\gtrsim0.2$~\cite{Almumin_2022}
\begin{equation}
    E(Q)_{\omega_0=0}\simeq \frac{5}{2}(\frac{\pi m_\phi^3 \phi_0^2}{3})^{1/5} (2N+1)^{3/5}Q^{4/5}\,,
    \label{53}
\end{equation}
\begin{equation}
         E(Q)_{\omega_0\neq0}\simeq \omega_0 Q+(2N+1)(\frac{\pi}{3})^{1/3} \frac{3^{2/3}\sqrt{m_\phi^2-\omega_0^2}}{2(\omega_0/\phi_0)^{2/3}}Q^{2/3}\,.
         \label{54}
\end{equation}
This expression can be used to demonstrate that it is energetically more favorable for an excited Q-balls to decay into a ground state Q-ball by emitting scalars instead of breaking into a number of smaller Q-balls as shown in Ref.~\cite{Almumin_2022} for global excited Q-balls.

\section{Summary and conclusion\label{sum}}
The success of describing unexcited gauged Q-balls properties via the mapping relation in terms of the simpler global Q-ball has inspired this article where we extend the method to discuss excited gauged Q-balls. The extension allows us to numerically obtain solutions for all the excited states and analytically derive expressions that approximates the properties of these states in the thin wall limit. It was shown that regardless of the excitation level, properties of gauged Q-balls are described by a single radius like length scale just like unexcited gauged Q-balls.

This is demonstrated in the paper by numerically producing excited gauged Q-ball profiles via the finite element method allowing us to compute exact properties such as radius, charge and energy of each excited state. Using the mapping relation, we derived an analytical expression that approximates the upper bound on the number of excited states a gauged Q-balls can posses. In the thin wall limit, analytical limits on the maximal radius are estimated in certain regions of the parameter space and approximate expressions of the charge and energy are produced for all excited states in terms of the excitation level. We discuss the possibility of having excited gauged Q-balls with larger lifetimes in certain parameter space where the energy gap between the excited states and ground state are shown to be smaller. We also show that analytical results are in good agreement with the exact results produced numerically in the large radius limit and close by discussing the unstable nature of these excited gauged Q-balls.

In conclusion, the mapping relation proved to be successful in describing gauged Q-balls in terms of global Q-balls beyond the ground state. The map produced an analytical approximation limit on the number of excited states a gauged Q-ball can acquire. Excited gauged Q-balls can be produced numerically using the finite element method, and analytical approximation are reliable in predicting the properties of these states in the thin-wall limit by specifying a single radius. Some regions of the parameter space admit excited gauged Q-ball with longer lifetime, which might be worth further investigation for potential phenomenological implications if the lifetime is large enough by properly studying the decay rates. Finding higher order corrections to small radii could also improve the analytical approximations.

\section*{Acknowledgements}
We are grateful for Arvind Rajaraman, Julian Heeck, Christopher B. Verhaaren and Rebecca Riley for giving us access to the code they developed to map gauged Q-balls~\cite{Heeck_2021map}, and for their insightful comments on the manuscript. The author is also thankful for Arvind Rajaraman and Michael Ratz academic guidance. The research of the author was supported by Kuwait University.

\bibliographystyle{utcaps_mod}
\bibliography{BIB}

\end{document}